\documentclass[twoside,twocolumn,9pt]{article}
\usepackage{extsizes}
\usepackage[super,sort&compress,comma]{natbib} 
\usepackage[version=3]{mhchem}
\usepackage[left=1.5cm, right=1.5cm, top=1.785cm, bottom=2.0cm]{geometry}
\usepackage{balance}
\usepackage{mathptmx}
\usepackage{sectsty}
\usepackage{lastpage}
\usepackage[format=plain,justification=justified,singlelinecheck=false,font={stretch=1.125,small,sf},labelfont=bf,labelsep=space]{caption}
\usepackage{float}
\usepackage{fancyhdr}
\usepackage{fnpos}
\usepackage[english]{babel}
\addto{\captionsenglish}{
  
}
\usepackage{booktabs}
\usepackage{dcolumn}
\usepackage{units}
\usepackage{array}
\usepackage{droidsans}
\usepackage{charter}
\usepackage[T1]{fontenc}
\usepackage[usenames,dvipsnames]{xcolor}
\usepackage{setspace}
\usepackage{nicefrac, xfrac}
\usepackage[compact]{titlesec}
\usepackage{hyperref}

\usepackage{subfig}

\newcommand*{\citen}[1]{
  \begingroup
    \romannumeral-`\x 
    \setcitestyle{numbers}
    \cite{#1}
  \endgroup   
}
\newcommand{\subfigimg}[3][,]{
  \setbox1=\hbox{\includegraphics[#1]{#3}}
  \leavevmode\rlap{\usebox1}
  \rlap{\hspace*{10pt}\raisebox{\dimexpr\ht1-2\baselineskip}{#2}}
  \phantom{\usebox1}
}
\usepackage{epstopdf}

\definecolor{cream}{RGB}{222,217,201}

\newcolumntype{f}{>{$}l<{$}}
\newcolumntype{n}{l}
\newcolumntype{N}{>{\scriptsize}l}
\newcolumntype{v}[1]{>{\raggedright\hspace{0pt}}p{#1}}
\newcolumntype{V}[1]{>{\scriptsize\raggedright\hspace{0pt}}p{#1}}

\newcolumntype{B}[1]{>{\boldmath\DC@{.}{,}{#1}}l<{\DC@end}}
\newcolumntype{d}[1]{>{\DC@{.}{,}{#1}}l<{\DC@end}}
\newcolumntype{i}[1]{>{\DC@{.}{,}{#1}\mathnormal\bgroup}l<{\egroup\DC@end}}
\newcolumntype{s}[1]{>{\DC@{.}{,}{#1}\mathsf\bgroup}l<{\egroup\DC@end}}

\newcolumntype{R}[1]{%
  >{\begin{turn}{90}\begin{minipage}{#1}\scriptsize\raggedright\hspace{0pt}}l%
  <{\end{minipage}\end{turn}}%
}

\newcolumntype{x}{>{\scriptsize\raggedright\hspace{0pt}}X}
\makeatother

\begin{document}

\pagestyle{fancy}
\thispagestyle{plain}
\fancypagestyle{plain}{

\renewcommand{\headrulewidth}{0pt}
}

\makeFNbottom
\makeatletter
\renewcommand\LARGE{\@setfontsize\LARGE{15pt}{17}}
\renewcommand\Large{\@setfontsize\Large{12pt}{14}}
\renewcommand\large{\@setfontsize\large{10pt}{12}}
\renewcommand\footnotesize{\@setfontsize\footnotesize{7pt}{10}}
\makeatother

\renewcommand{\thefootnote}{\fnsymbol{footnote}}
\renewcommand\footnoterule{\vspace*{1pt}%
\color{cream}\hrule width 3.5in height 0.4pt \color{black}\vspace*{5pt}} 
\setcounter{secnumdepth}{5}

\makeatletter 
\renewcommand\@biblabel[1]{#1}            
\renewcommand\@makefntext[1]%
{\noindent\makebox[0pt][r]{\@thefnmark\,}#1}
\makeatother 
\renewcommand{\figurename}{\small{Fig.}~}
\sectionfont{\sffamily\Large}
\subsectionfont{\normalsize}
\subsubsectionfont{\bf}
\setstretch{1.125} 
\setlength{\skip\footins}{0.8cm}
\setlength{\footnotesep}{0.25cm}
\setlength{\jot}{10pt}
\titlespacing*{\section}{0pt}{4pt}{4pt}
\titlespacing*{\subsection}{0pt}{15pt}{1pt}

\fancyhead{}

\twocolumn[

\begin{center}
    \LARGE{\textbf{The role of thermalisation in hot carrier cooling dynamics}} 

 \vspace{2ex} 
    \large{T. Faber,\textit{$^{\dag}$} L. Filipovic,\textit{$^\ddag$} and L.J.A. Koster\textit{$^{\dag\ast}$}}\\
 \vspace{1cm} 
    \textbf{Abstract} 
\end{center}

\normalsize{The hot carrier solar cell (HCSC) concept has been proposed to
overcome the Shockley Queisser limit of a single p-n junction solar
cell by harvesting carriers before they have lost their surplus energy.
A promising family of materials for these purposes is metal halide
perovskites (MHP). MHPs have experimentally shown very long cooling times, the key
requirement of a HCSC. By using Ensemble Monte Carlo
(EMC) simulations we shed light on why cooling times are found to be extended for these materials. In this manuscript, we concentrate on the role of thermalisation in the cooling process. We specify the role of electron-phonon and electron-electron interactions in thermalisation and cooling, while furthermore showing how these processes depends on several relevant material parameters, such as the dielectric constant and the effective mass. Finally, we quantify how thermalisation can also act as a cooling mechanism via the cold background effect. Here, we stress the importance of a low degree of background doping in order to achieve the observed extended cooling times. This work provides insights into the ongoing discussion on cooling times in MHPs. In addition our results are an important addition to the debate on whether or not tin perovskites are suitable candidates for
HCSCs.}\vspace{1cm}] 

\renewcommand*\rmdefault{bch}\normalfont\upshape
\rmfamily
\section*{}
\vspace{-1cm}

\footnotetext{\textit{$^{\dag}$~Zernike Institute for Advanced
Materials, University of Groningen, Nijenborgh 4, 9747 AG Groningen, The Netherlands.}}
\footnotetext{\textit{$^{\ddag}$~CDL for Multi-Scale Process Modeling of Semiconductor Devices and Sensors at the Institute for Microelectronics, TU Wien,
Gusshausstrasse 27-29
1040 Vienna, Austria.}}

\frenchspacing

\section{Introduction}

Hot carrier solar cells (HCSC) have been proposed to overcome the Shockley Queisser (SQ) limit,\cite{KahmannLoi} the theoretical efficiency limit for a single junction solar cell.\cite{Shockley} \cite{wurfel2000physik} The rationale behind the SQ limit is threefold. One photon only creates a single electron-hole pair, the solar cell is under unconcentrated illumination, and all excess energy is transferred to the lattice via thermal relaxation. By harvesting charge carriers before they have lost their surplus energy in the form of heat, the theoretical power conversion efficiency (PCE) could be boosted up to 66 \% under 1 sun illumination. \cite{Ross} However, as carriers typically cool down within 100 fs,\cite{Nelson} the realization of a HCSC has appeared to be a daunting task and until the present day no working device has been presented. 

More recently, a resurgence of interest has been initiated by reports of metal halide perovskites (MHP) showing unusually long cooling times (see Table 1). \cite{Li} MHPs have already been in the center of the attention for quite some time for their excellent opto-electronic properties, while being solution processable and cost-efficient.\cite{Wang} Intriguingly, the experimental reports on slow cooling in MHPs show significant variation in their results, triggering a scientific debate on the physical explanations behind this phenomena. \cite{Yang_HPB, Niesner, Frost, Fu, Fang}  As slow cooling is eminent for a HCSC to work, the discussion calls for understanding and further research on the topic.

\begin{figure}[ht]
\centering
  \includegraphics[height=5cm]{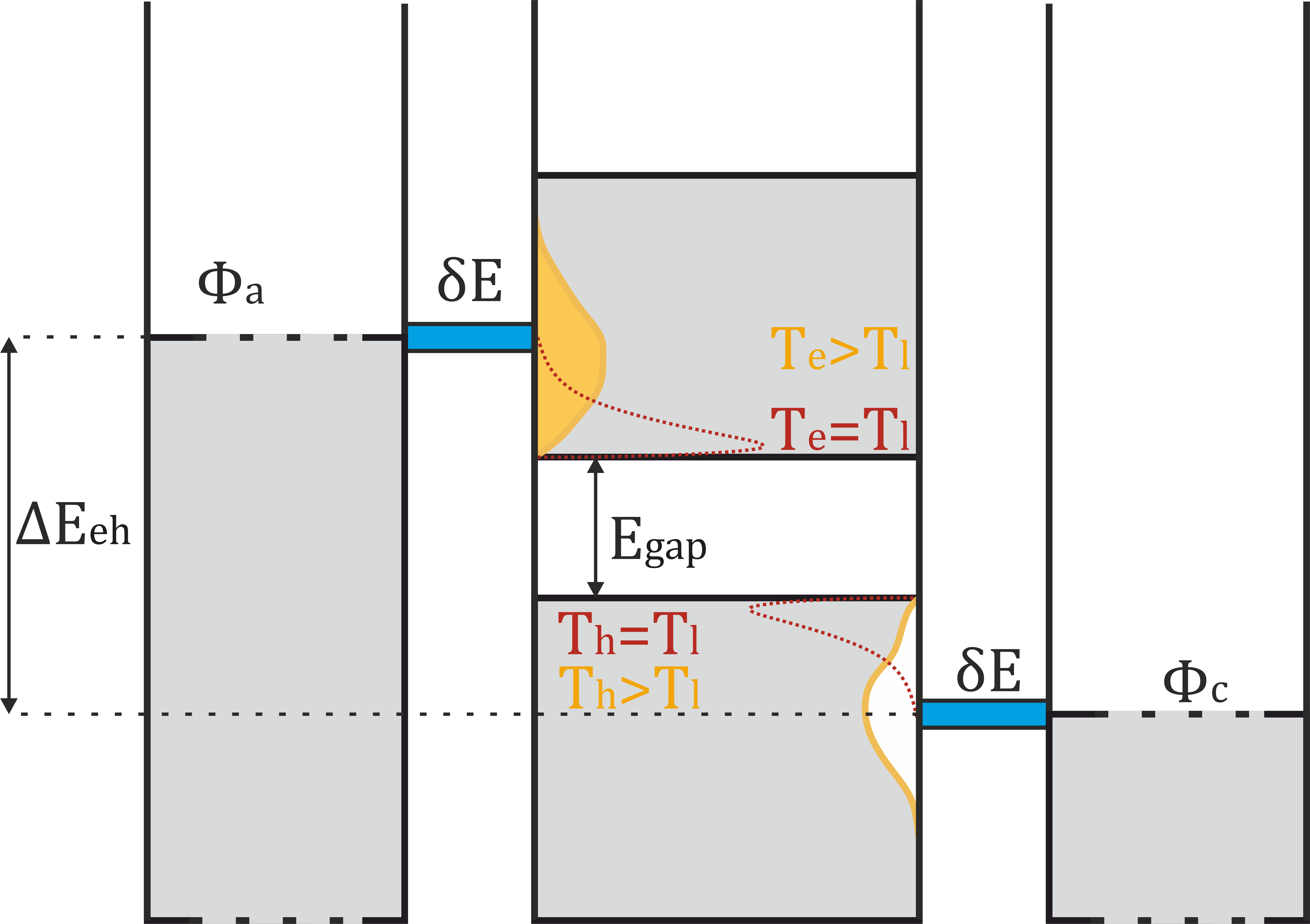}
  \caption{The design of a hot carrier solar cell consists of an active layer wedged in between two energy selective contacts (ESC). These ESCs are aligned with the hot carrier distributions (yellow). These distributions are at elevated temperatures $T_e$ and $T_h$, compared to the regular, cooled down, distributions (red) at lattice temperature $T_l$. The corresponding energy difference between the two quasi-Fermi levels is therefore larger resulting in a higher open circuit voltage.}
  \label{fgr:HCSC}
\end{figure}

The design of HCSCs depicted in Fig. 1 depends on several factors. \cite{Koning_2010} The most important is an absorber layer which presents extended relaxation times, i.e., carriers must lose their energy as slowly as possible, since only then does it becomes possible to extract them while they are still hot. Obviously the absorber layer must also possess features of ‘normal’ PV materials, such as a reasonably high mobility and broad spectral absorption.\cite{KahmannLoi}  

A second critical aspect of HCSCs is that carriers are extracted to the contacts with the help of energy selective contacts (ESC).\cite{Koning_2010, Limpert} The energy level of these contacts are aligned with the hot carrier distributions at temperatures $T_e $ and $T_h$ (see Fig. 1), enabling the extraction carriers while they are still hot. ESCs are small bandwidth ($\sim$ 0.1 eV), large bandgap materials, which transport carriers \textit{isentropically} to the contacts, maintaining the enlarged energy difference ($\Delta E_{eh}$) between the quasi-Fermi levels and hence resulting in an enhanced open circuit voltage. \cite{Limpert} 

\begin{figure*}[ht!]
 \includegraphics[width=\textwidth]{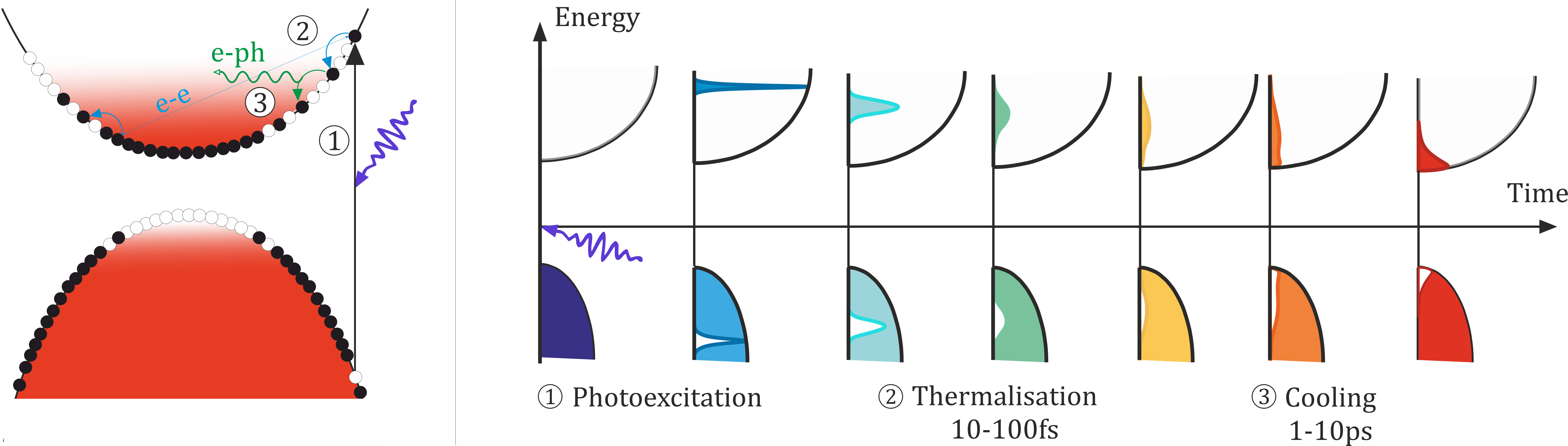}

  \caption{General picture of hot carrier dynamics in a semiconductor. Carriers are photo-excited at step 1, forming non-equilibrium distributions in the respective bands. Carriers subsequently thermalise into Fermi-Dirac distributions characterized by a temperature at step 2. For MHPs, thermalisation is dominated by carrier-carrier interactions. Finally, at step 3, carriers cool down towards the lattice temperature via carrier-phonon interactions, before they eventually recombine.}
  \label{fgr:HC_dynamics}
\end{figure*}

After photo-excitation (step 1, Fig. 2) the carriers form non-equilibrium distributions in their respective bands.\cite{Koning_2010} These distributions quickly equilibrate into Fermi Dirac distributions, defined by respective temperatures $T_e$ and $T_h$ (step 2, Fig. 2). This process is defined as carrier thermalisation, as one can now assign a temperature to the distribution.\footnote{Often the relaxation timescale is referred to as the thermalisation time,\cite{Conibeer_thermalisation} referring to the thermalisation losses which occur when the carrier cools down. However this is physically not accurate. Thermalisation is the process of acquiring a temperature, hence we therefore stick to the definition stated above, following Ref. [\citen{Nelson}]} Carrier thermalisation can occur via carrier-carrier, carrier-phonon and carrier-impurity scattering.\cite{Richter} In the case of MHPs, carrier-carrier scattering is assumed to be the dominating process, due to their relatively low optical dielectric constant, resulting in weaker Coulomb screening.\cite{Richter} Thermalisation occurs on a fs to a sub-ps timescale, and is expected to occur much faster than the relaxation process.\cite{Rota} It is important to note that thermalisation is a continuous process, and keeps on influencing the cooling dynamics even after the carriers cool further down via carrier-phonon interactions (step 3, Fig. 2). In polar semiconductors, such as MHPs, relaxation predominantly occurs via the Fröhlich interaction with longitudinal optical (LO) phonons.\cite{Frohlich, Science-hotcarriers}

Understanding carrier thermalisation is of central importance for the realization of HCSCs. The reason for this is twofold:
\begin{enumerate}
    \item Fast thermalisation is an essential ingredient in HCSCs for successful operation under steady state conditions.\cite{Conibeer_2010} As carriers are extracted from the absorber by the ESCs, the carrier distribution is perturbed at the energy level of the ESC, and must be rethermalised as quickly as possible in order for optimal carrier extraction to be maintained.\cite{Conibeer_2010} It is essential to understand how one should tune the material, in order to achieve fast thermalisation times, while keeping cooling times extended. 
    \item In ultrafast spectroscopy experiments, thermalisation via carrier-carrier interactions also plays an important role in the cooling dynamics in HCSCs.\cite{Richter, Ulatowski} As was pointed out by Richter et al.,\cite{Richter} and emphasized by Ulatowski et al.,\cite{Ulatowski} an increase in the degree of background doping could play a detrimental role on the cooling time. Carriers would lose their surplus energy much faster as they now thermalise not only among other non-equilibrium carriers, but also with a background of carriers, resulting into a much lower $T_e$ / $T_h$. Essentially, cooling occurs via thermalisation. Insight in this effect could particularly be important for the analysis on the suitability of tin based perovskites for HCSCs. Tin based perovskites have  shown very long cooling times;\cite{Fang} however, the oxidation of \ce{Sn^{2+}} to \ce{Sn^{4+}} inherently produces a large background ensemble of cold holes,\cite{tin1, tin2} which could be detrimental for the hot carrier dynamics.\cite{Kahmann_purity, Ulatowski}
    
\end{enumerate}


\begin{table*}[ht]
\renewcommand*{\arraystretch}{1.5}
\centering
\caption{Selection of observed experimental cooling times in metal halide perovskites\cite{Li}}
\begin{tabular}[t]{l>{\raggedright}p{0.1\linewidth}>{\raggedright}p{0.15\linewidth}>{\raggedright}p{0.12\linewidth}>{\raggedright\arraybackslash}p{0.12\linewidth}>{\raggedright\arraybackslash}p{0.14\linewidth}}
\toprule
 &\textbf{Relaxation times to \\600 K (ps)} &Carrier density \\ ($\rho$ in 10$^{17}$ cm$^{-3}$)  & Dielectric constants \\ ($\epsilon_{\infty}$ / $\epsilon_{0}$) & Electron effective mass ($m^*$) & Phonon frequency  ($\omega_0$ / 2$\pi$ in THz)\\
\midrule
\ce{GaAs}&0.1&1&7.9\cite{GaAs-diel}/12\cite{GaAs-diel}&0.0067\cite{GaAs-mass} & 8.15\cite{GaAs-freq}\\

\ce{MAPbI_3}&0.6&5.2&4.5\cite{Brivio_2013} / 25.7\cite{effmassMAPbI3}&0.15\cite{effmassMAPbI3}&2.25\cite{Frost} \\

 &60&60& & \\

&1.1&64& & \\

\ce{MAPbBr_3}&0.8&150& 6.7\cite{Zhao} / 25\cite{Zhao} &  0.27\cite{Filippo}&4.47\cite{Frost}  \\

 &200&0.7& & \\
\ce{CsPbI_3}  & 2 &  7 & 6.1\cite{Frost}/ 18.1\cite{Frost} &0.17\cite{Ponce} &2.57\cite{Frost} \\

&10&70& & \\
\ce{FAPbI_3}  & 40 &  11 & 6.6\cite{FaPbI3-eps} / - & 0.1\cite{FaPbI3-effmass} & 2.25 \cite{FAPbI3-freq}\\

\ce{FASnI_3}  &   1000 &  60 & 3.58\cite{FaSnI3-diel} / - &  0.17\cite{FaSnI3-diel}& - \\
\bottomrule

\end{tabular}
\end{table*}%


In this contribution we investigate both roles mentioned above, in order to understand the role that thermalisation plays in the hot carrier (HC) dynamics of a potential HCSC. We do this by making use of an Ensemble Monte Carlo (EMC) simulation framework.\cite{Ferry, Jacoboni, Jacoboni2, Hess, Hockney, Tomizawa} The EMC technique is a proven numerical method, when dealing with transport in semiconductors,\cite{Kosina} especially when it comes to far-from-equilibrium transport.\cite{Ferry} Mainly due to its adjustability, and the full description of $x$ and $k$ space, when treating scattering, the EMC method is an excellent tool to investigate scattering mechanisms and their respective interplay. Over the years, it has been frequently used to deal with hot carrier problems,\cite{Ferry, GaAs_EMC, Lugli_EMC, Duncan_EMC} for example in studies of hot carrier trends in metal oxide semiconductor field effect transistors,\cite{Duncan_EMC} or more closely related to our investigation, when modeling the electron-hole interactions on the ultrafast relaxation of hot photoexcited carriers in GaAs.\cite{GaAs_EMC} However in the field of MHPs, EMC has only been used in order to quantify the effect of polaron formation on the mobility of charge carriers.\cite{Walker}

The EMC fully takes into account both scattering and transport processes, and is therefore the necessary next step in accurately describing the cooling process of hot carriers in MHPs.\cite{Science-hotcarriers} Our model consists of carrier-carrier scattering and carrier-LO-phonon interactions, as thermalisation is governed by the former and relaxation by the latter. In this manuscript, we focus on using MHPs as an absorber material in order to contribute to the recent developments on reported HC cooling of MHPs; however we note that our discussion has broader implications.

We begin by visualizing the HC dynamics in a \ce{MAPbI3}-like system and report similar relaxation times as those documented experimentally.  Our study is finally able to bridge hot carrier behavior with theory. An interpretation of the results obtained with HC experiments relies heavily on HC theory. Therefore, a direct simulation and visualisation on how exactly theoretical parameters impact the process could be helpful for further interpretation of observations and, ultimately, can serve to inform an improved design. We subsequently examine thermalisation in detail. First, focusing on reason 1 given above, we address the question of how one can achieve fast thermalisation times, while keeping relaxation times extended. We show the trends for the thermalization and relaxation times for several material parameters, mainly the effective mass and dielectric permittivity. We choose these parameters specifically as they both impact both carrier-carrier and Fröhlich interactions, causing the resulting effect on the thermalisation and cooling process to be non-trivial. We find that a small effective mass, and thereby a sharp energy band, is desired for both fast thermalisation and slow cooling, and show that the optical dielectric constant plays a more delicate role, as weaker screening results in faster thermalisation. However could also increase relaxation rates. Second, we shift our investigation towards reason 2. Here we show and quantify the effect of thermalisation with a cold-background ensemble on the cooling time, concluding that the effect is significant while being highly density dependent. Our work provides insights on the current discussion on cooling times of MHPs, and the obtained theoretical understanding could be helpful when designing to design a working HCSC.

\section{Methods}
For our simulations we used an Ensemble Monte Carlo code ViennaEMC.\cite{vienna-EMC} The Ensemble Monte Carlo method\cite{Jacoboni} solves the Boltzmann transport equation without a priori assumptions regarding the form of the distribution function, and offers the implementation of complex band structures,\cite{Jacoboni2} making it a highly accurate method for studying the time-evolution of a statistical ensemble.\cite{phd-EMC} The Monte Carlo method models the motion of an ensemble of electrons as a sequence of randomly generated free flights, interrupted by randomly selected scattering events.\cite{phd-EMC} For the generation of random numbers, a long period  ($2 \times  10^{18}$) L'Ecuyer random number generator\cite{Ecuyer} was used with Bays-Durham shuffle and added safeguards. In the EMC, an entire ensemble of particles is simulated one after the other for the period of one timestep, tracking both $x$ and $k$ space coordinates. The method is semi-classical as the the free-flight is modelled classically, while the scattering rates are computed quantum mechanically via Fermi's Golden Rule.\cite{Jacoboni} The scattering rates are energy dependent, pre-computed, and stored at intervals of 0.4 meV in look-up tables. Our model consists of scattering via LO-phonon interactions, described by Fröhlich interactions, and carrier-carrier interactions via Coulomb interactions. The LO-phonon scattering rate for both absorption and emission is given by:\cite{Jacoboni}
\begin{align}
    P_{e,LO} (E) = \frac{e^2 \sqrt{m^*} \omega_0}{\sqrt{2}\hbar} \bigg(\frac{1}{\epsilon_{\infty}} - \frac{1}{\epsilon_0}\bigg)\textbf{F}(E, E') \bigg[N_{op} + \frac{1}{2} \mp \frac{1}{2}\bigg],
\end{align} where $e$ is the elementary charge, $m^*$ is the effective mass, $\omega_0$ is the typical phonon frequency, $\epsilon_{\infty}, \epsilon_0$ are the optical and static part of the dielectric constant. The phonon number is given by $N_{op}$ for absorption (-) and $N_{op} + 1$ for emission (+). Finally, $E$ is the energy and $\textbf{F}(E,E')$ is a function of the energy $E$ before, and after $E'$ interaction with a phonon, where $E' = E \pm \hbar \omega_{op} $ for absorption (-) and emission (+) respectively .  

The carrier-carrier interactions were modelled classically using the Fast-Multipole Method (FMM) using the ScalFMM package.\cite{ScalFMM} The FMM handles particles via direct compution at close range, while clusters of distant particles are grouped together and approximated by the use of a multipole expansion.\cite{rokhlin} More information on how the FMM is implemented in detail can be found in the original paper of Rohklin.\cite{rokhlin} Since the FMM models the carrier-carrier interactions classically, the electric field is recomputed after every timestep, before each free-flight phase commences. the close-range interactions in the FMM are sensitive to divergences,\cite{vasileska} therefore a ${(r+a)}$ kernel was used with a = 1 nm. The Coulomb potential can then be represented as
\begin{equation}
    V_c(r) = \frac{q}{4\pi \epsilon (r + a)}.
\end{equation} The dielectric constant $\epsilon$ used in the carrier-carrier interactions was taken to be equal to the optical part of the dielectric constant $\epsilon_{\infty}$, which also returns in the LO-phonon scattering rate. In our simulations the multipole expansion was solved up to the second order, resulting into a computation error of $10^{-2}$ for the potential, without the introduction of an external field. Ensembles of 10000 to 100000 particles, consisting of 50 \% holes and 50 \% electrons, were simulated in a 100 nm to 1 $\mu$m box, which was periodically repeated in all three directions 384 times. A parabolic band structure approximation was used defined by an effective mass. The particle dynamics are then described by
\begin{equation}
     v = \frac{1}{\hbar} \nabla_k E(k) = \frac{\hbar k}{m^*}, 
\end{equation}
and
\begin{equation}
       r(t) = r(0) + v t + \frac{q E(r)}{2 m^*} t^2,
\end{equation}
where $r$ is the position, $k$ is the momentum, $\hbar$ is the reduced Planck constant, $\textbf{E}(k)$ is the electric field, $m^*$ the effective mass in terms of the electron mass $m_e$, and t is the time. These equations were solved for each timestep (0.1 fs) for the entirety of the simulation time.

\section{Results and Discussion}
\begin{figure*}[ht!]
    \centering
    \subfloat[\centering $\rho$ = $10^{16}$ cm$^{-3}$ ]{{\includegraphics[width=0.5\textwidth, height=5.3cm ]{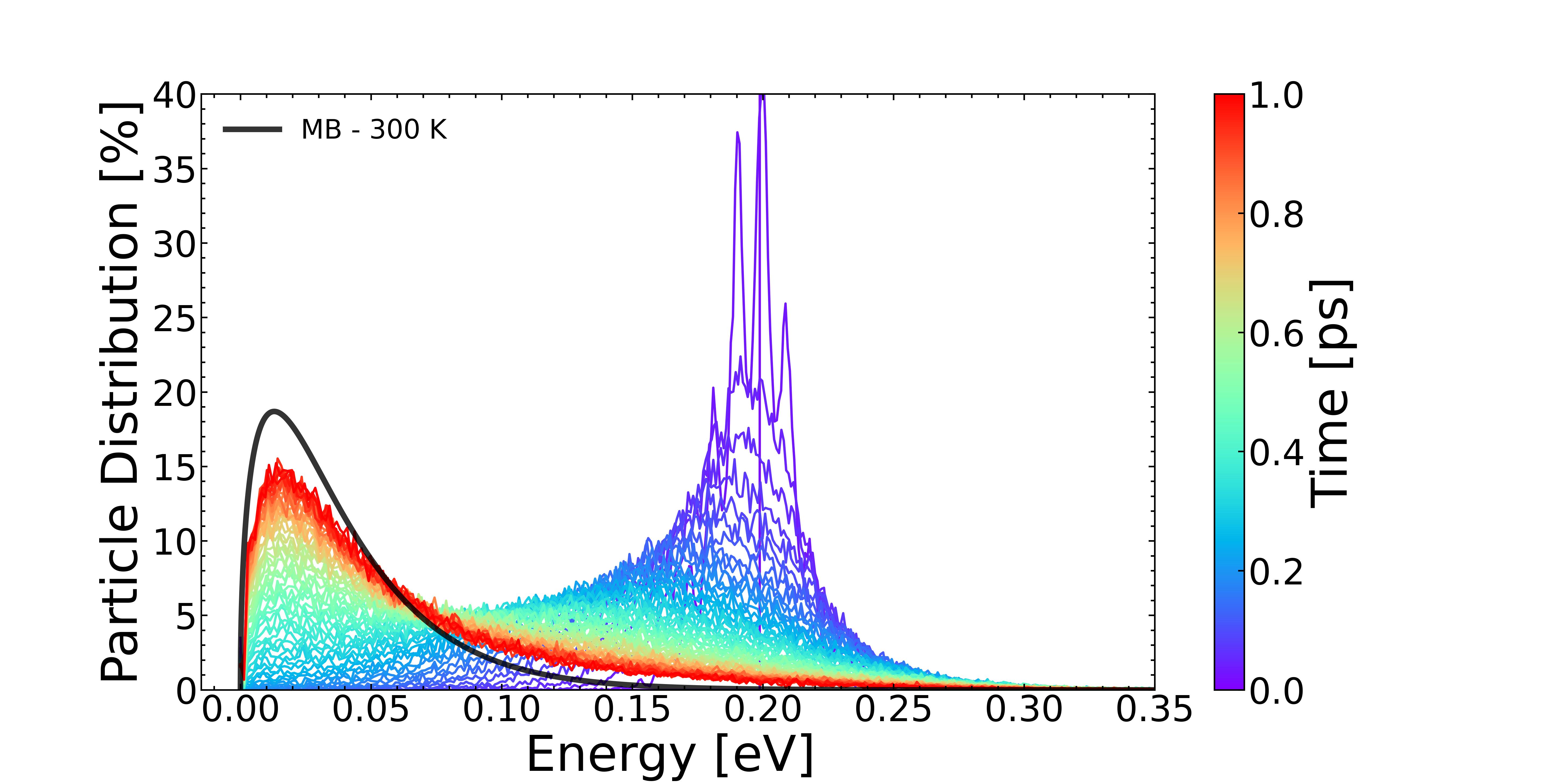} }}%
    \subfloat[\centering $\rho$ = $10^{16}$ cm$^{-3}$ ]{{\includegraphics[width=0.5 \textwidth, height= 5cm]{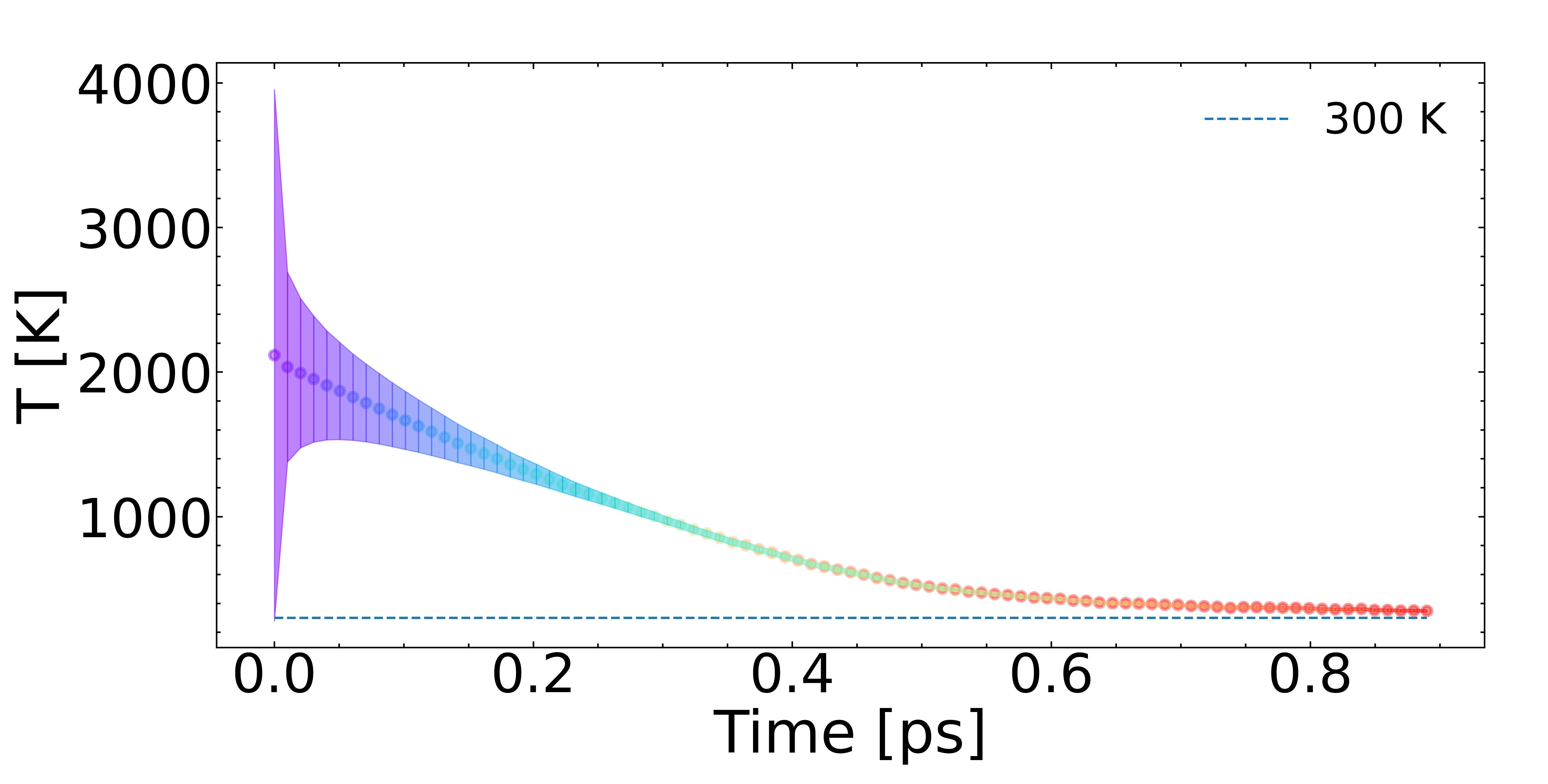} }}%
    \qquad
    \subfloat[\centering $\rho$ = $10^{18}$ cm$^{-3}$
    ]{{\includegraphics[width=0.5 \textwidth, height= 5.3cm]{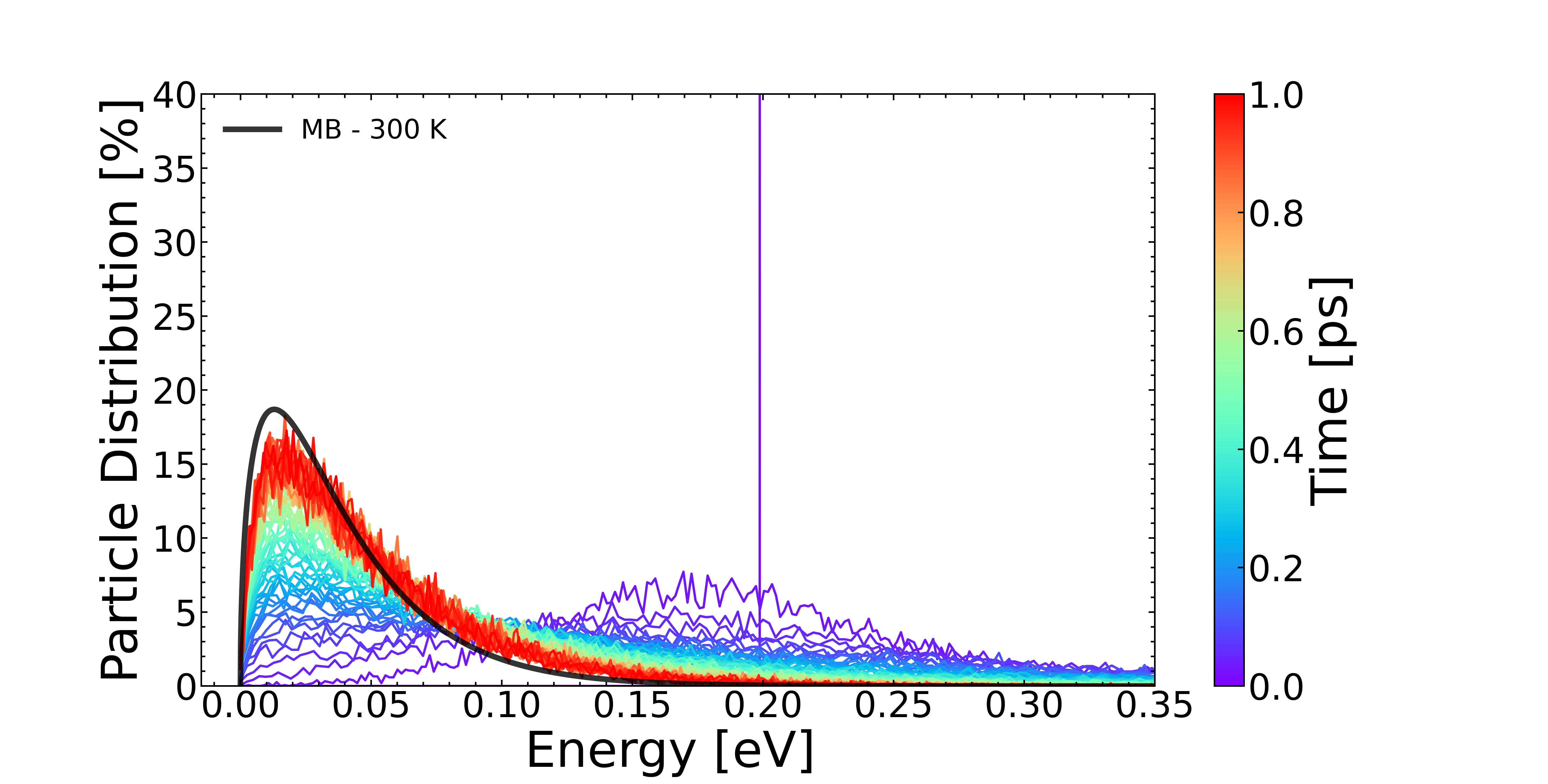} }}%
    \subfloat[\centering $\rho$ = $10^{18}$ cm$^{-3}$ ]{{\includegraphics[width=0.5 \textwidth, , height=5cm]{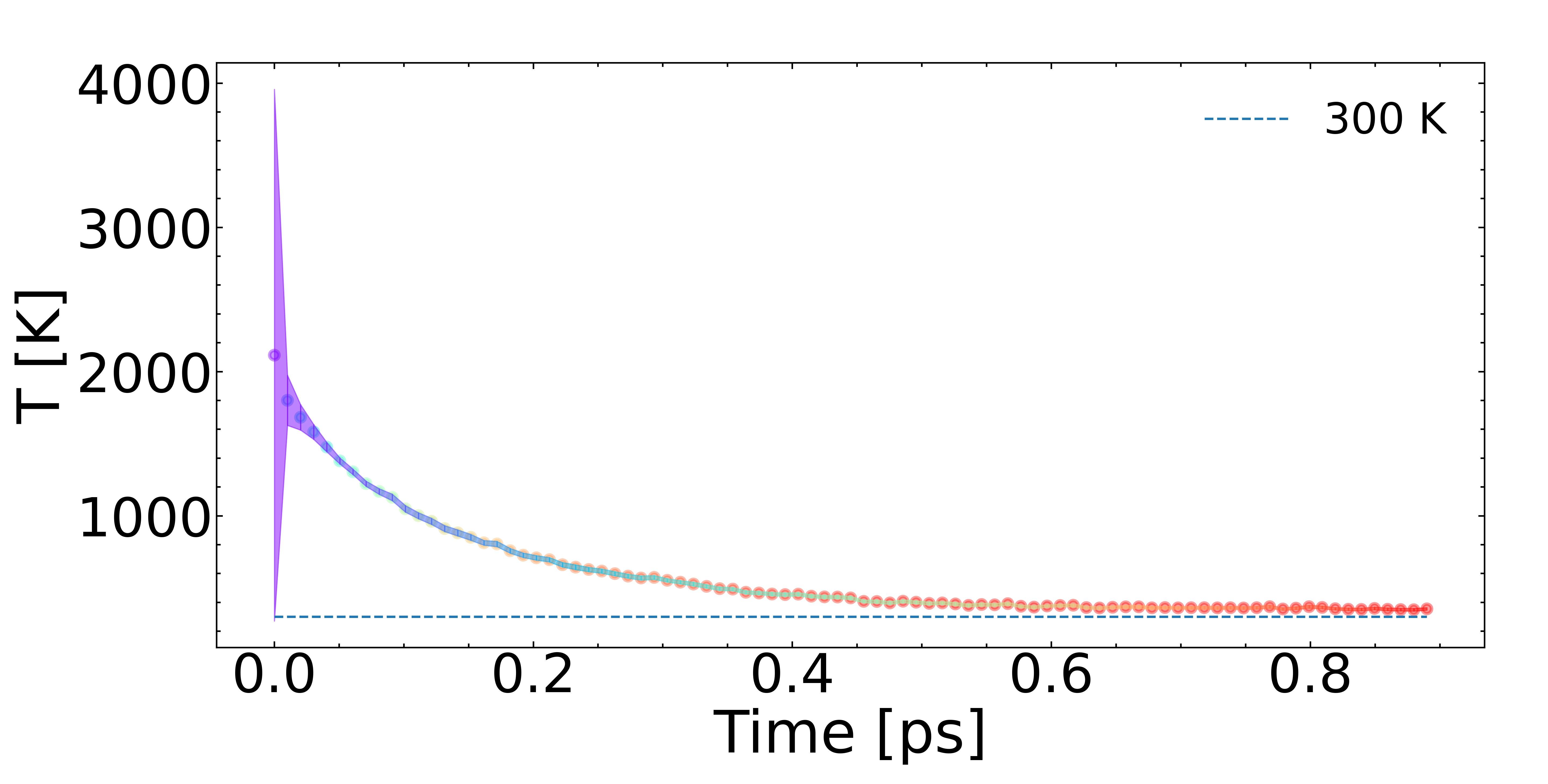} }}
    \caption{The cooling dynamics of \ce{MAPbI3} ($m^* = 0.15 $, $\epsilon_{\infty} = 4.5$, $\epsilon_{0} = 25.7$, and $\omega_0$ = 2.25 THz) with a), c) the energy distribution for $\rho$ = $10^{16}$ cm$^{-3}$, and $10^{18}$ cm$^{-3}$ and b), d) the temperature evolution over time for $\rho$ = $10^{16}$ cm$^{-3}$, and $10^{18}$ cm$^{-3}$. }%
    \label{fig:energytempdistributions}%
    
\end{figure*}
Our investigation starts with the visualisation of the hot carrier cooling dynamics for metal halide perovskites, particularly within a typical \ce{MAPbI3} like system. Our motivation for the focus on a \ce{MAPbI3} like system, is that most studies of hot carrier cooling have been focused on this perovskite. We define a \ce{MAPbI3} like system by the effective mass $m^*$, the dielectric constant $\epsilon_{\infty}$, and the typical phonon frequency $\omega_0$. The principle aim is to answer the question whether or not the cooling process of \ce{MAPbI3} can accurately be described by using EMC simulations. Our starting situation consists of an ensemble of carriers, both electrons and holes, in thermal equilibrium, interacting via carrier-phonon coupling and carrier-carrier interactions, at a temperature of 300 K. The carriers carry an effective mass of $m^* = 0.15 $ both for electrons and holes.\cite{effmassMAPbI3} The optic and static part of the dielectric constants are $\epsilon_{\infty} = 4.5$,\cite{Brivio_2013} and $\epsilon_{0} = 25.7$\cite{effmassMAPbI3} respectively, and the typical phonon frequency for \ce{MAPbI3} is given by $\nicefrac{\omega_0}{2\pi}$ = 2.25 THz.\cite{Frost} The simulations are initiated, at time $t = 0$, with an energy pulse of 0.2 eV, exciting the entire ensemble. Here, we give all the particles a kinetic energy of 0.2 eV, and randomize their momenta in all three spatial directions. Subsequently we track the position and momenta of all particles over a period of 1 ps. The temperature of the system can be obtained by fitting the kinetic energy distribution to a Maxwell-Boltzmann distribution, defined by a temperature. Moreover, the fitting procedure also yields the degree of thermalisation. By tracking how accurately the energy distribution fits to a Maxwell-Boltzmann distribution over time, we are able to measure the degree of the thermalisation over time.  In Fig. 3 we display the results for both carrier densities of $\rho$ = $10^{16}$ cm$^{-3}$, and $\rho$ = $10^{18}$ cm$^{-3}$, where the former corresponds to a carrier-phonon dominated regime, and the latter to a carrier-carrier dominated one. With higher particle densities, carrier-carrier interactions are expected to become more and more dominant, as the Coulomb interaction scales with $\rho^\frac{1}{3}$.  Figures 3a, c describe the evolution of the energy distribution, while Fig.\ 3b, d show the temperature of the system over time. In Fig.\ 3b, d, the fitting error is plotted around the best-fit temperature as a measure of the thermalisation. 

From Fig.\ 3, one can observe the different stages of the cooling process as described above. At $\rho$ = $10^{16}$ cm$^{-3}$, Fig. 3b, we observe the system to thermalise in within about $\sim$ 200 fs, while cooling down from approximately 2000 K to 300 K in around 1 ps.  At $\rho$ = $10^{18}$ cm$^{-3}$, Fig. 3d, thermalisation is much faster $\sim$ 40 fs, however cooling time is on the same scale. We note that our result at $\rho$ = $10^{16}$ cm$^{-3}$ is in good agreement with experimental values found for \ce{MAPbI3}, as Yang et al. report values of 0.6 ps for cooling down from 1500 K to 600 K at a density of $10^{17}$ cm$^{-3}$.\cite{Yang_HPB} However, cooling times reported by   Yang et al. significantly increase for higher densities, while our cooling times do not. The authors attribute the extended cooling times to a hot-phonon bottleneck effect.\cite{hotphbottle, hotphbottle2} The hot phonon bottleneck effect is a term used to encompass all phenomena which lead to a increased probability of phonon re-absorption, due to a non-equilibrium phonon population which comes into play above a critical density.\cite{hotphbottle, hotphbottle2} Yang et al. found that the critical density is at a density of $\rho \sim 5 \cdot 10^{17}$ cm$^{-3}$.\cite{Yang_HPB} The hot-phonon bottleneck effect is a second order effect, where phonon-phonon interactions play a crucial role. Our model does not include phonon-phonon interactions, since the focus of this paper is the interplay between carrier-phonon and carrier-carrier interactions. As such, a discussion this difference is beyond the scope of the presented model.

\begin{figure}[t!]
    \centering
    \subfigimg[width=0.44\textwidth, height=5cm]{a)   }{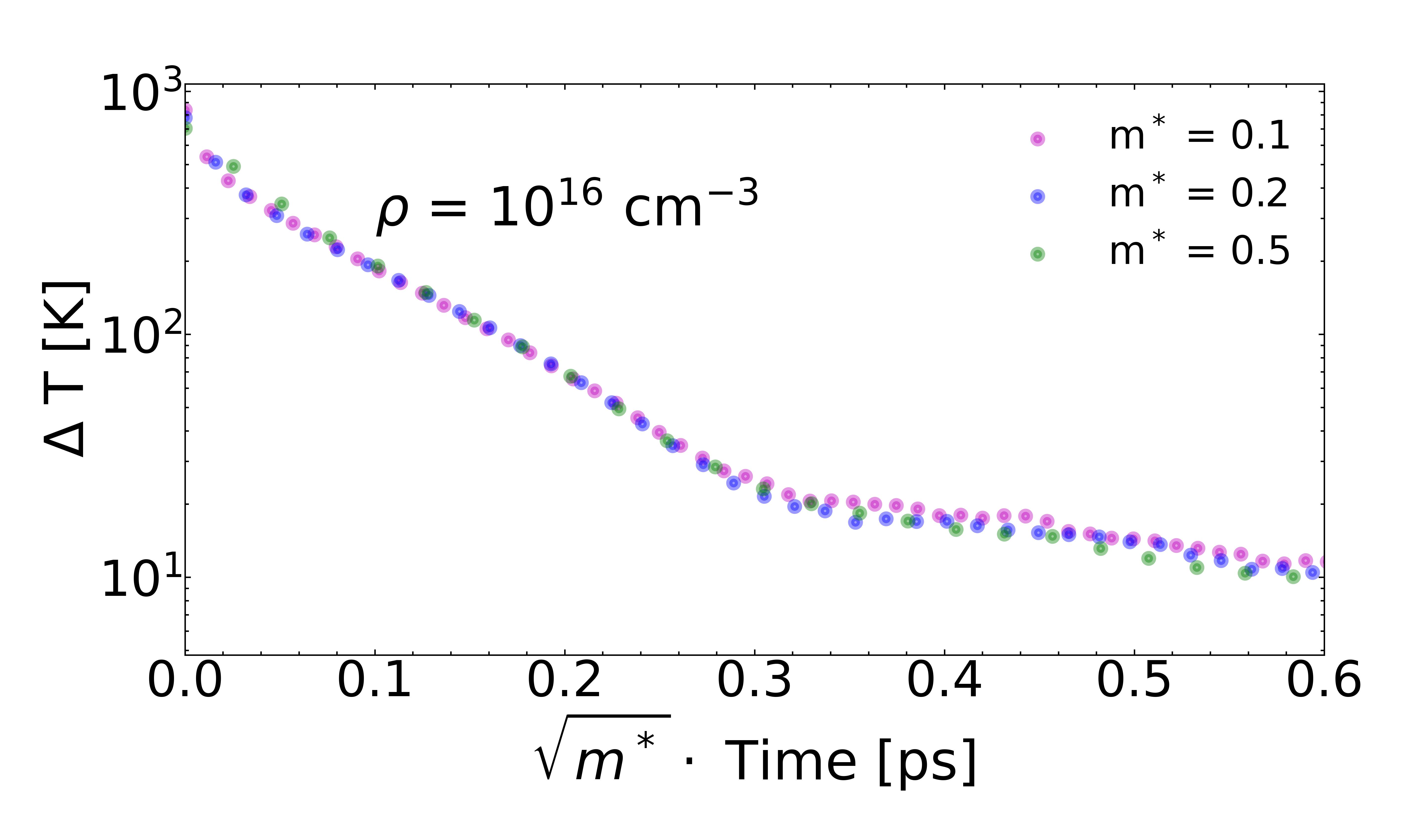} %
    \\
    \subfigimg[width=0.44\textwidth, height=5cm]{b)}{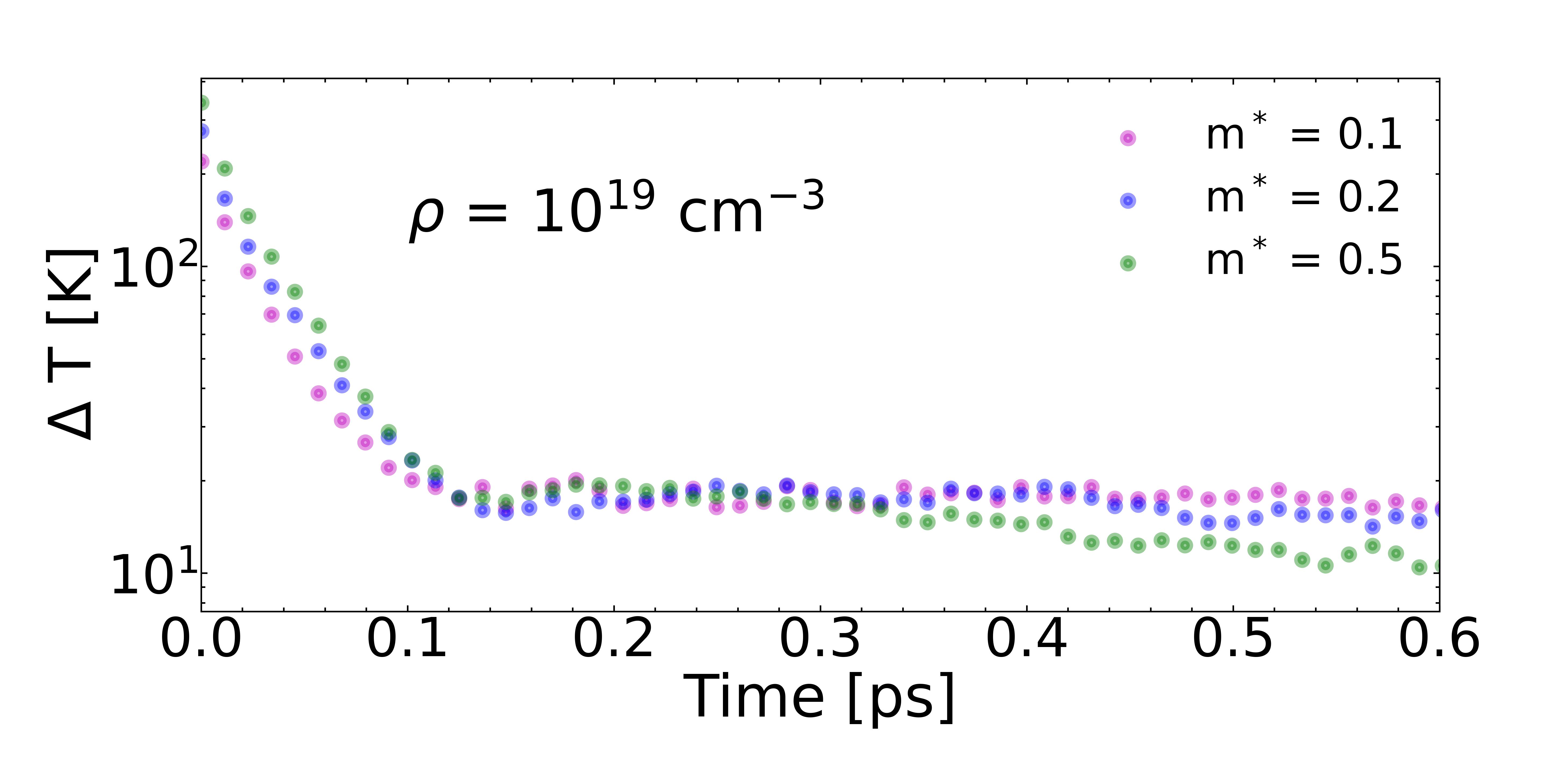}%
    \caption{Thermalisation times for different values of the effective mass $m^*$ for different densities a) $\rho$ = $10^{16}$ cm$^{-3}$ and b) $\rho$ = $10^{19}$ cm$^{-3}$.}
    \label{fig:effmass}%
\end{figure}

Unlike for cooling times, we do observe density dependence in the thermalisation time. A comparison between Fig. 3a and Fig. 3c makes it evident how much faster the higher density distribution takes on a smoothed form. The discrete nature of the energy spectrum shown in Fig. 3a at early times indicates that timescales of carrier-phonon and carrier-carrier interactions for these systems at this density are comparable. The more continuous distribution at higher density (Fig. 3c) points to carrier-carrier interactions being the much more dominant process behind the thermalisation process. In this limit, the thermalisation time depends on the number of scattering events between carriers per unit of time, which is obviously increased for higher carrier densities. This is also consistent with experimental results at this density for perovskites.\cite{Richter} 

 Our result on the density dependence on the thermalisation time invites us to return to the question of how a material should be tuned in order for it to have fast thermalisation, while preserving slow cooling. Even though thermalisation is dominated by carrier-carrier interactions, it is not completely determined by it. As one can see in Fig. 3a and Fig. 3b thermalisation times are about 2 to 3 times shorter than cooling times, indicating also the relevance of carrier-phonon interactions on the thermalisation process.  In perovskite-like systems, thermalisation is dominated by carrier-carrier interactions, and cooling by LO-phonon coupling. In Fig.\ 4 and Fig. 5 we present the impact of the effective mass and the dielectric constant on the thermalisation time, for different densities. Both parameters are present in the calculation of the carrier-phonon and the carrier-carrier interactions. This creates the opportunity for an in-depth analysis on the interplay between both interactions governing the different processes.

At time $t = 0$ the entire ensemble was excited to an energy of 0.2 eV, and the temperature of the system was subsequently tracked and noted down every 10 fs for a period of 1 ps. In Fig.\ 4 and Fig. 5, $\Delta T$ represents the deviation of the current energy distribution from that of a Maxwell-Boltzmann distribution. As defined above, when $\Delta T \rightarrow 0$, the state becomes thermalised. Plotting this over time, yields a measure of the thermalisation time. Simulations were performed for different values of the effective mass ($m^*  =$ 0.1, 0.2, 0.5) and the dielectric constant ($\epsilon_{\infty} =$ 5, 10, 15) at particle densities ($\rho$ = $10^{16}$ cm$^{-3}$, $10^{19}$ cm$^{-3}$). Here we go to even higher densities, as we intend to observe what happens in the limit where carrier-carrier interactions are strongly dominant. The values of the material parameters span over a range corresponding to perovskite-like systems, i.e., polar semiconductors characterized by relatively low effective masses ($ m^* < 0.5 $) and a low degree of screening ($\epsilon_{\infty} \leq 15 $). 

The impact of the effective mass is given in Fig. 4a, b. At a density of $\rho = 10^{16}$ cm$^{-3}$ (Fig. 4a), we observe that thermalisation takes place faster for a larger effective mass. Due to the low particle density, carrier-phonon interactions might be the dominant interaction mechanism. By looking at the LO-phonon scattering rate we obtain a $\sqrt{m^*}$ dependency. As we plot this data against time $\cdot \sqrt{m^*}$, we obtain an excellent agreement (Fig. 4a), confirming our assumption. The observation of carrier-phonon interactions dominating the process can be supported by noting the relatively long timescales of the thermalisation time in this limit.

At a high particle density $\rho$ = $10^{19}$ cm$^{-3}$(Fig. 4b), the thermalisation time is found to be independent of the effective mass. More accurately, the phonon $\sqrt{m^*}$ dependency is compensated by a $\frac{1}{\sqrt{m^*}}$ dependency coming from the electron-electron scattering rate $k_{ee}$.\cite{delfatti}.  We observe thermalisation times to be faster in this limit, supporting the hypothesis of a regime change. Further support can be obtained by looking at Fig. S1, where we show the dependency of the effective mass on the thermalisation time for a system with only carrier-carrier interactions to follow a $\frac{1}{\sqrt{m^*}}$ trend. This is in agreement with the established theory on the electron-electron scattering rate.\cite{delfatti}

We further analyze our results for the dielectric constant in Fig 5. Generally, it can be observed that the thermalisation time scales inversely with the dielectric constant. As particles experience less screening, they thermalise faster. We find a $\epsilon_{\infty}^{-\frac{3}{2}}$ fit for the data at a low density (Fig. 5a) and a $\epsilon_{\infty}^{-2}$ fit at a high density (Fig. 5b). Both dependencies fit the data very well, however an insightful nuance regarding interaction regimes can be distilled from the two different fits. The $\epsilon_{\infty}^{-\frac{3}{2}}$ fit corresponds to a regime where the carrier-phonon interactions are balanced by the carrier-carrier interactions. The dependencies on $\epsilon_{\infty}$ for the LO-phonon coupling and Coulomb interaction are $\frac{1}{\epsilon_{\infty}}$, noted in eq. (1), and $\epsilon_{\infty}^{-2}$\cite{delfatti} respectively. A $\epsilon_{\infty}^{-\frac{3}{2}}$ dependency (Fig. 5a) would indicate a regime where both interactions are exactly balanced. By looking more closely at the high density data (Fig. 5b), one can observe that the overlap between the yellow line ($\epsilon_{\infty} = 10$) and the red line ($\epsilon_{\infty} = 15$) is actually more precise. This perfect fit for a $\epsilon_{\infty}^{-2}$ dependency suggests a carrier-carrier dominated regime, following similar reasoning as above.

\begin{figure}[t!]
    \centering
    \subfigimg[width=0.44\textwidth, height=5cm]{a)}{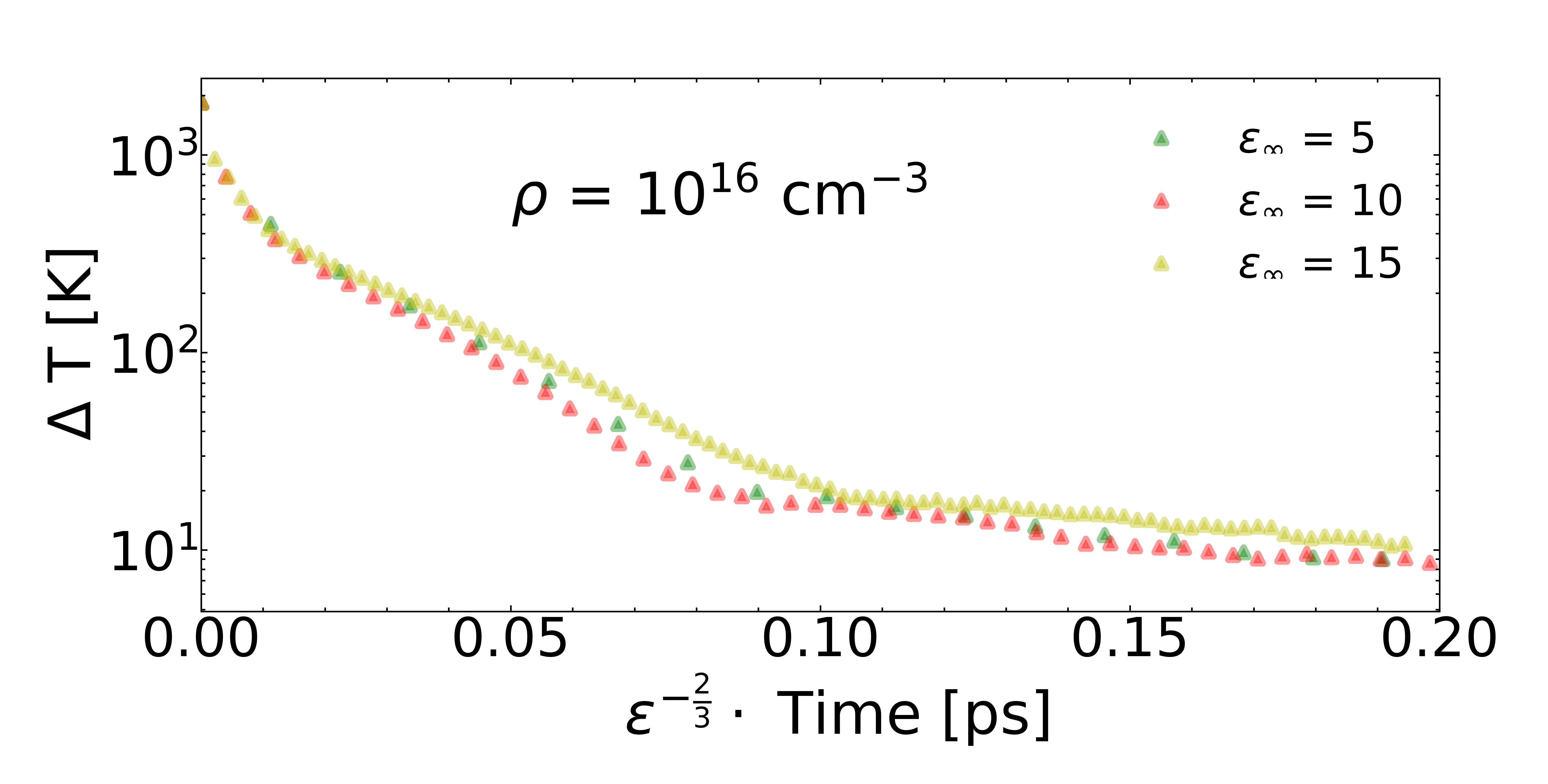} %
    \\
    \subfigimg[width=0.44\textwidth, height=5cm]{b)}{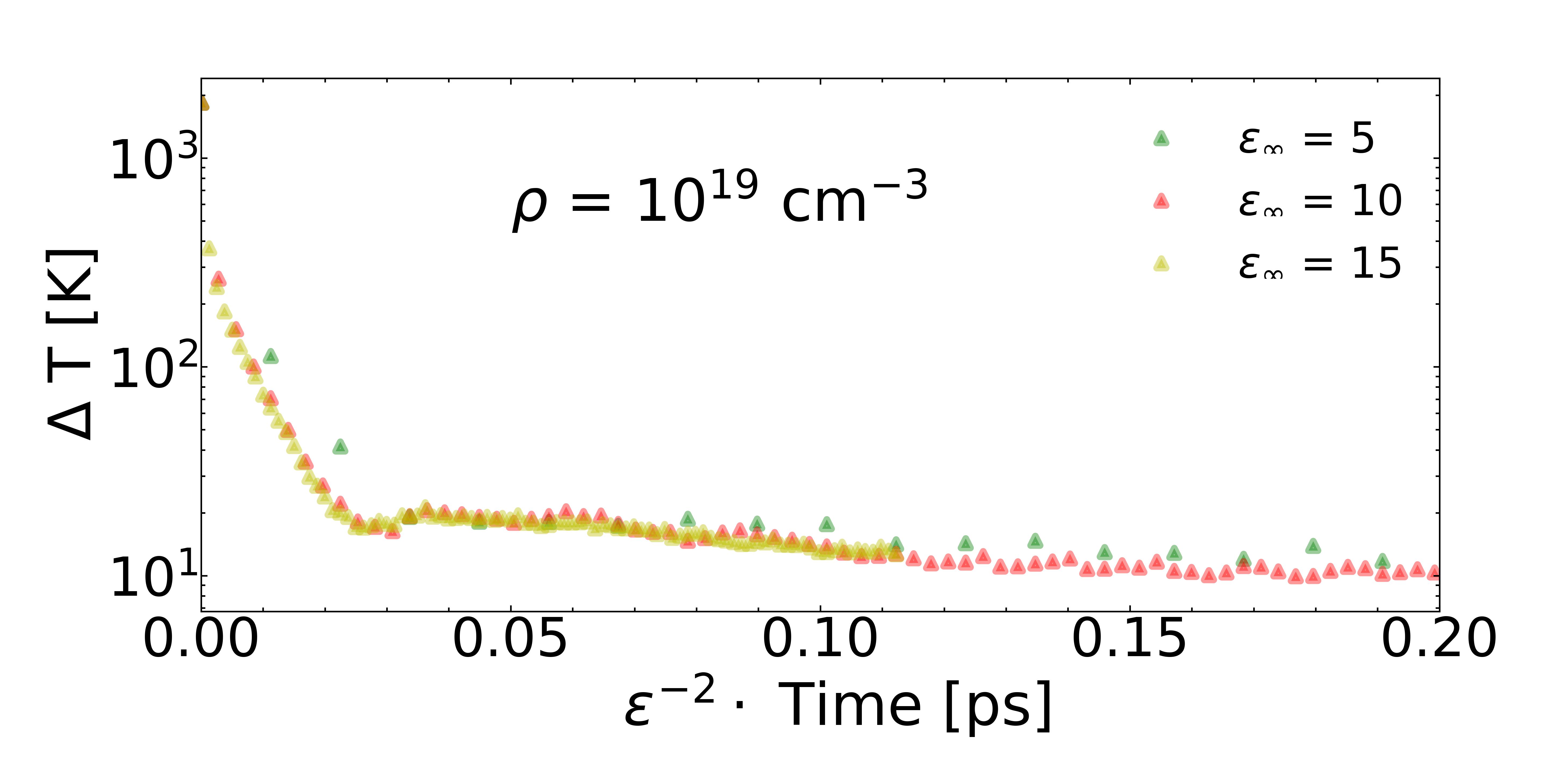} 
    \caption{Thermalisation times for different values of the dielectric constant $\epsilon_{\infty}$ for different densities a) $\rho$ = $10^{16}$ and b) $\rho$ = $10^{19}$ cm$^{-3}$.}
    \label{fig:eps}%
\end{figure}

Intuitively, it could be argued that for a higher degree of screening, a drift away from a carrier-carrier dominated regime would be expected, as the Coulomb interaction scales inversely proportional with the dielectric of the material. However, also the carrier-phonon interactions are diminished by a larger $\epsilon_{\infty}$. Moreover, as the Fröhlich interaction is proportional to the difference between the static and the optic part of the dielectric constant, noted in eq. (1), the $\frac{1}{\epsilon_{\infty}}$ proportionality is only valid in the limit where $\epsilon_{\infty} \ll \epsilon_{0}$. When $\epsilon_{0} = 25$ in these simulations, the values for $\epsilon_{\infty}$ used approach the asymptotic limit. As a matter of fact, an increase in $\epsilon_{\infty}$ from 5 to 15 results into a decrease in $P_{e-LO}$ by a factor of 9. One can conclude that the increased Coulomb screening is eclipsed by a decrease in carrier-phonon interaction, resulting into a carrier-carrier dominated regime for higher values of the dielectric constant. 

The LO-phonon interaction strength plays an important part in distinguishing between different regimes. Therefore, also a quantity such as the phonon frequency $\omega_0$ impacts these results, as a large phonon frequency results in a stronger phonon coupling. Our results show how significant carrier-phonon interactions can be for thermalisation in low particle density perovskite systems. In the high particle density regime, which is generally the operating regime for a HCSC,  the influence of carrier-phonon interactions is less significant. In this regime, a small the effective mass is desired as it slows down cooling, does not impact thermalisation negatively, and enhances the mobility. A low optic dielectric constant $\epsilon_{\infty}$ is desired for fast thermalisation. However, in combination with a high static dielectric constant $\epsilon_{0}$ it could result in fast cooling. It would be informative to see how these results would alter in a system where phonon-phonon interactions are incorporated, as in this case perhaps strong LO-phonon coupling is desired for a fast population of non-equilibrium phonon modes.\cite{Yang_HPB}

\begin{figure}[t!]
    \centering
    \subfigimg[width=0.47\textwidth]{a)}{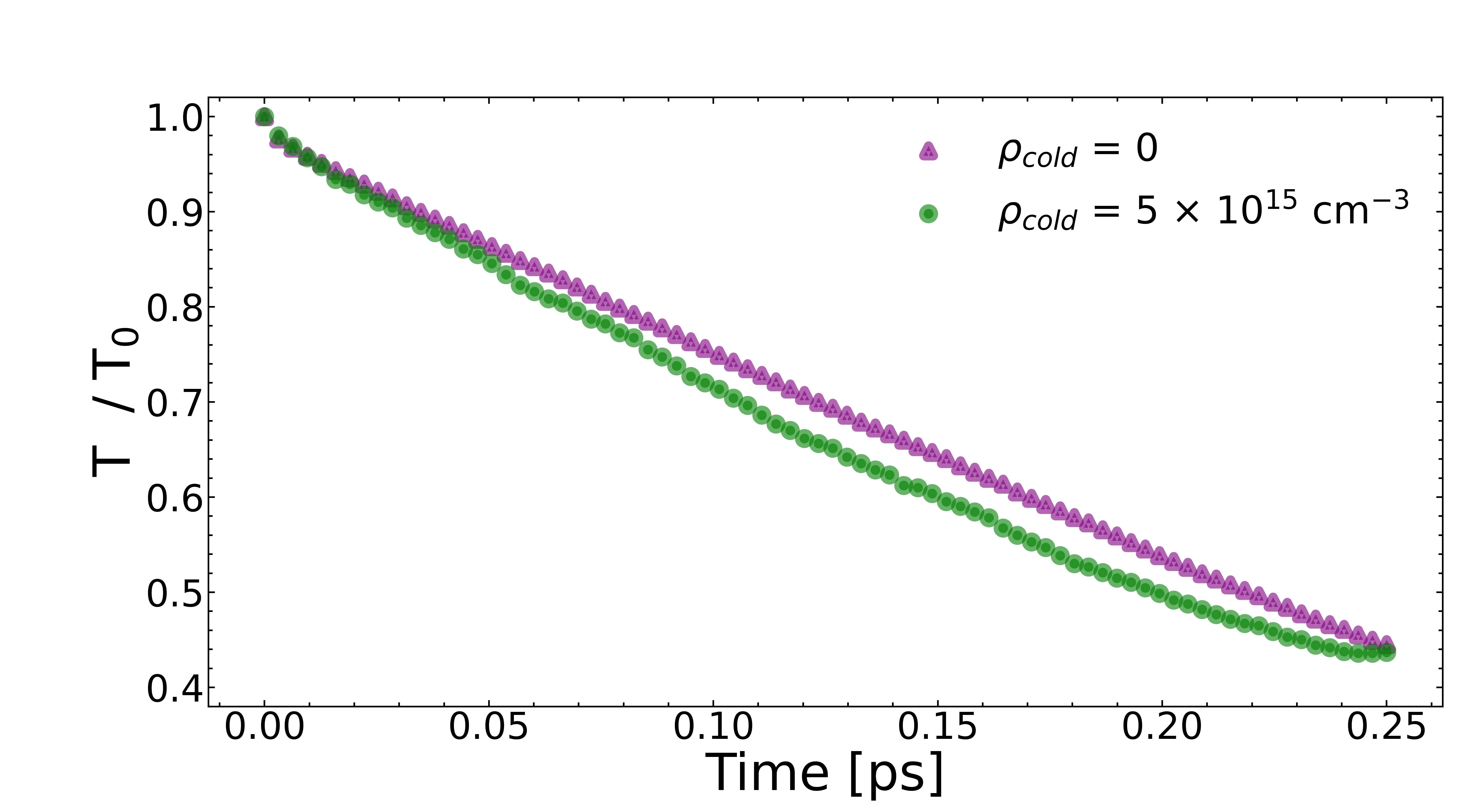} %
    \\
    \subfigimg[width=0.47\textwidth]{b)}{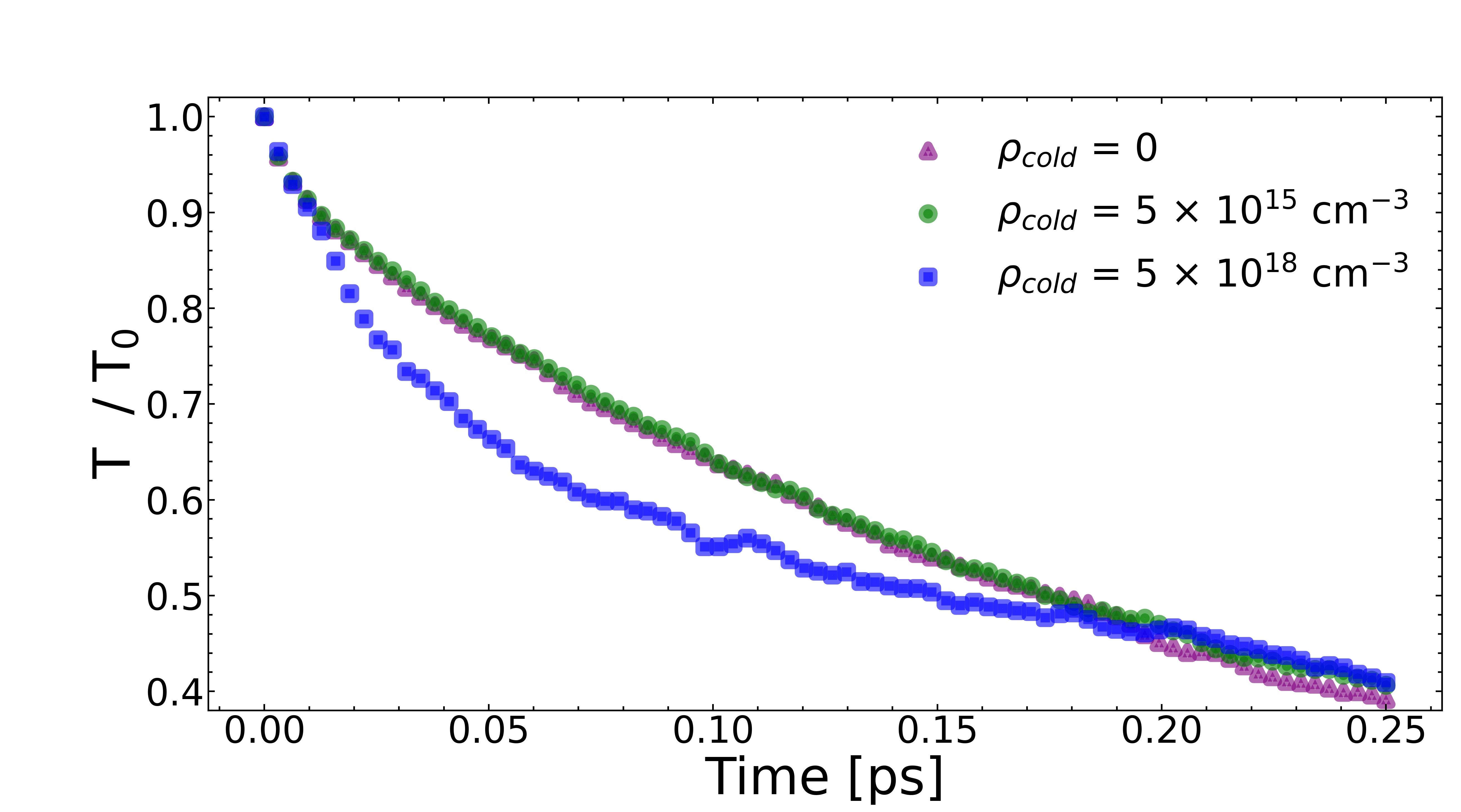}%
    \label{fig:CBE}%
    \caption{Normalised temperature over time after initiating an energy pulse for a perovskite-like system with different background densities $\rho_{cold}$. Simulations are performed for different excitation densities  a) $\rho_{photo} = 10^{16}$ cm$^{-3}$, b) $\rho_{photo} = 10^{19}$ cm$^{-3}$.}
\end{figure}

Finally, we discuss the effect a cold background ensemble of carriers can have on the cooling dynamics of newly excited carriers. We have performed simulations on a thermal ensemble of carriers at 300K of a perovskite-like system. We start by introducing an energy pulse of 0.2 eV, and observing the cold background effect by varying the background density $\rho_{cold}$. The density of the ensemble which is excited by the energy pulse is labeled as $\rho_{photo}$. We track the normalized temperature $T / T_0$ over time, in order to observe whether or not there is a different cooling gradient in the initial thermalisation phase. We show the cold background (CB) effect in Fig. 6 for two different density regimes. We only plot the first 0.25 ps because the effect stops as soon as thermalisation is completed.

Our results indicate a strong cold background effect at a high background density. Fig. 6a shows our results for low carrier density $\rho_{photo} = 10^{16}$ cm$^{-3}$. We do not observe any significant difference in the cooling slope, when we vary the background density from 0 cm$^{-3}$ to $5 \times 10^{15}$ cm$^{-3}$. A further increase of the background density to $5 \times 10^{18}$ cm$^{-3}$ would give meaningless results, as the energy pulse in a system like this would be comparable to a perturbation. At a high carrier density, $\rho_{photo} = 10^{19}$ cm$^{-3}$ we note a strong effect at earlier times. When the background density is increased from $5 \times 10^{15}$ cm$^{-3}$ to $5 \times 10^{18}$ cm$^{-3}$, we observe a steeper cooling slope for the first 0.1 ps to 0.2 ps. We can attribute the different results at different densities to the fact that the CB-effect is governed by carrier-carrier interactions, which are highly density dependent in nature as we have already observed in the thermalisation times of \ce{MAPbI_3} (Fig. 4b, and Fig. 4d). Likewise, other parameters, such as the dielectric constant and the effective mass. which impact the carrier-carrier scattering rate, enhance this effect. However, as they also impact the phonon cooling rate, it is more difficult to quantify their impact on the CB-effect. 

Our findings confirm several important experimental results in understanding the large differences found for relaxation times in tin-perovskites. The self-doping property of tin via the oxidation of \ce{Sn^{2+}} to \ce{Sn^{4+}} could result into the formation of a large cold background ensemble of holes.\cite{tin1, tin2} The degree of background doping has been put forward to address the wide range of observed cooling times.\cite{Kahmann_purity}\cite{Ulatowski}  Our results show accelerated cooling during the first scattering phase for high background density systems, which confirms the hypothesis used by Kahmann et al.\cite{Kahmann_purity} and Ulatowksi et al,\cite{Ulatowski} to explain the shorter cooling times for tin based perovskites. 

The strong density dependence for this effect makes it possible to limit its consequences by choosing an operating regime where carrier-carrier interactions do not dominate the thermalisation process. However, as fast thermalisation is also a key ingredient for the successful operation of a HCSC, this could become quite difficult. We do note that a cold background ensemble only forms if the operation of the HCSC is not optimal. In theory, carriers should be extracted before they reach the bottom of the energy band, and fast thermalisation is an essential component to repopulate the extraction level.

As a final note, it is important to mention the topic of polarons. Polaron formation has also been identified as a separate process in the cooling dynamics in MHPs, and has been put forward as a potential mechanism behind the slow cooling times.\cite{Zhu_polarons, Frost_hotpolarons, Niesner, Evans_polarons, Joshi_polarons, Burgos-Caminal} The incorporation of polarons into the EMC model would be a relevant, though complicated addition. Polaron formation could impact both thermalisation and cooling, as the effective mass of carriers is altered.\cite{Emin} A detailed description of the cooling dynamics for a polaronic carrier landscape would be a fruitful next step, but lies beyond the scope of this manuscript.

\section{Conclusions}
In this contribution we used ensemble Monte Carlo simulations to investigate the role of thermalisation on carrier cooling dynamics for perovskite-like systems. We have visualized the thermalisation and cooling process of \ce{MAPbI3} for different densities. Secondly we have shown how thermalisation times depend on different material parameters, and we have extracted trends showing how to optimize the thermalisation time. Our work puts emphasis on the fact that thermalisation and cooling are not two separated stages, but interconnected processes. By studying their interplay, we have qualified and quantified the impact of carrier-phonon interactions on the thermalisation process. 

We can conclude that for HCSC purposes a small carrier effective mass is desired. Our analysis on the dielectric constant showed that a low dielectric constant $\epsilon_{\infty}$ would enhance thermalisation; However, could also accelerate carrier cooling. Finally, we have quantified how thermalisation can act as a cooling mechanism via the cold-background effect. Here, we assessed that the effect is highly density dependent and only starts playing a role at high carrier densities and a high background density $\rho = 10^{19}$ cm$^{-3}$. These findings help explain the large differences reported in relaxation times for tin perovskites.

\section*{Conflicts of interest}
There are no conflicts to declare.

\section*{Acknowledgements}
The authors would like to thank the Center for Information Technology of the University of Groningen for their support and for providing access to the Peregrine high performance computing cluster, the Zernike Institute for Advanced Materials for funding. The financial support by the Austrian Federal Ministry of Labour and Economy and the National Foundation for Research, Technology and Development and the Christian Doppler Research Association is gratefully acknowledged. This work was supported in part by the Austrian Research Promotion Agency FFG (Bridge Young Scientists) under Project 878662 ”Process-Aware Structure Emulation for Device-Technology Co-Optimization”. Finally the authors would like to thank Federico Ferrari for the illustrations. 

\balance

\bibliography{main} 

\providecommand*{\mcitethebibliography}{\thebibliography}
\csname @ifundefined\endcsname{endmcitethebibliography}
{\let\endmcitethebibliography\endthebibliography}{}
\begin{mcitethebibliography}{62}
\providecommand*{\natexlab}[1]{#1}
\providecommand*{\mciteSetBstSublistMode}[1]{}
\providecommand*{\mciteSetBstMaxWidthForm}[2]{}
\providecommand*{\mciteBstWouldAddEndPuncttrue}
  {\def\EndOfBibitem{\unskip.}}
\providecommand*{\mciteBstWouldAddEndPunctfalse}
  {\let\EndOfBibitem\relax}
\providecommand*{\mciteSetBstMidEndSepPunct}[3]{}
\providecommand*{\mciteSetBstSublistLabelBeginEnd}[3]{}
\providecommand*{\EndOfBibitem}{}
\mciteSetBstSublistMode{f}
\mciteSetBstMaxWidthForm{subitem}
{(\emph{\alph{mcitesubitemcount}})}
\mciteSetBstSublistLabelBeginEnd{\mcitemaxwidthsubitemform\space}
{\relax}{\relax}

\bibitem[Kahmann and Loi(2019)]{KahmannLoi}
S.~Kahmann and M.~Loi, \emph{Journal of Materials Chemistry C}, 2019,
  \textbf{7}, 2471--2486\relax
\mciteBstWouldAddEndPuncttrue
\mciteSetBstMidEndSepPunct{\mcitedefaultmidpunct}
{\mcitedefaultendpunct}{\mcitedefaultseppunct}\relax
\EndOfBibitem
\bibitem[Shockley and Queisser(1961)]{Shockley}
W.~Shockley and H.~J. Queisser, \emph{Journal of Applied Physics}, 1961,
  \textbf{32}, 510--519\relax
\mciteBstWouldAddEndPuncttrue
\mciteSetBstMidEndSepPunct{\mcitedefaultmidpunct}
{\mcitedefaultendpunct}{\mcitedefaultseppunct}\relax
\EndOfBibitem
\bibitem[W{\"u}rfel(2000)]{wurfel2000physik}
P.~W{\"u}rfel, \emph{Physik der Solarzellen}, Spektrum Akademischer Verlag,
  2000\relax
\mciteBstWouldAddEndPuncttrue
\mciteSetBstMidEndSepPunct{\mcitedefaultmidpunct}
{\mcitedefaultendpunct}{\mcitedefaultseppunct}\relax
\EndOfBibitem
\bibitem[Ross and Nozik(1982)]{Ross}
R.~T. Ross and A.~J. Nozik, \emph{Journal of Applied Physics}, 1982,
  \textbf{53}, 3813--3818\relax
\mciteBstWouldAddEndPuncttrue
\mciteSetBstMidEndSepPunct{\mcitedefaultmidpunct}
{\mcitedefaultendpunct}{\mcitedefaultseppunct}\relax
\EndOfBibitem
\bibitem[Nelson(2003)]{Nelson}
J.~Nelson, \emph{The Physics Of Solar Cells}, World Scientific Publishing
  Company, 2003\relax
\mciteBstWouldAddEndPuncttrue
\mciteSetBstMidEndSepPunct{\mcitedefaultmidpunct}
{\mcitedefaultendpunct}{\mcitedefaultseppunct}\relax
\EndOfBibitem
\bibitem[Li \emph{et~al.}(2019)Li, Fu, Xu, and Sum]{Li}
M.~Li, J.~Fu, Q.~Xu and T.~C. Sum, \emph{Advanced Materials}, 2019,
  \textbf{31}, 1802486\relax
\mciteBstWouldAddEndPuncttrue
\mciteSetBstMidEndSepPunct{\mcitedefaultmidpunct}
{\mcitedefaultendpunct}{\mcitedefaultseppunct}\relax
\EndOfBibitem
\bibitem[Wang \emph{et~al.}(2019)Wang, Yang, Wu, Sanghadasa, and Priya]{Wang}
K.~Wang, D.~Yang, C.~Wu, M.~Sanghadasa and S.~Priya, \emph{Progress in
  Materials Science}, 2019, \textbf{106}, 100580\relax
\mciteBstWouldAddEndPuncttrue
\mciteSetBstMidEndSepPunct{\mcitedefaultmidpunct}
{\mcitedefaultendpunct}{\mcitedefaultseppunct}\relax
\EndOfBibitem
\bibitem[Yang \emph{et~al.}(2015)Yang, Ostrowski, France, Zhu, van~de Lagemaat,
  Luther, and Beard]{Yang_HPB}
Y.~Yang, D.~Ostrowski, R.~France, K.~Zhu, J.~van~de Lagemaat, J.~Luther and
  M.~Beard, \emph{Nature Photonics}, 2015, \textbf{10}, 53–59\relax
\mciteBstWouldAddEndPuncttrue
\mciteSetBstMidEndSepPunct{\mcitedefaultmidpunct}
{\mcitedefaultendpunct}{\mcitedefaultseppunct}\relax
\EndOfBibitem
\bibitem[Niesner \emph{et~al.}(2016)Niesner, Zhu, Miyata, Joshi, Evans,
  Kudisch, Trinh, Marks, and Zhu]{Niesner}
D.~Niesner, H.~Zhu, K.~Miyata, P.~P. Joshi, T.~J.~S. Evans, B.~J. Kudisch,
  M.~T. Trinh, M.~Marks and X.-Y. Zhu, \emph{Journal of the American Chemical
  Society}, 2016, \textbf{138}, 15717--15726\relax
\mciteBstWouldAddEndPuncttrue
\mciteSetBstMidEndSepPunct{\mcitedefaultmidpunct}
{\mcitedefaultendpunct}{\mcitedefaultseppunct}\relax
\EndOfBibitem
\bibitem[Frost(2017)]{Frost}
J.~Frost, \emph{Physical Review B}, 2017, \textbf{96}, 195202\relax
\mciteBstWouldAddEndPuncttrue
\mciteSetBstMidEndSepPunct{\mcitedefaultmidpunct}
{\mcitedefaultendpunct}{\mcitedefaultseppunct}\relax
\EndOfBibitem
\bibitem[Fu \emph{et~al.}(2017)Fu, Xu, and Han]{Fu}
J.~Fu, Q.~Xu and G.~Han, \emph{Nature Communications}, 2017, \textbf{8},
  1300\relax
\mciteBstWouldAddEndPuncttrue
\mciteSetBstMidEndSepPunct{\mcitedefaultmidpunct}
{\mcitedefaultendpunct}{\mcitedefaultseppunct}\relax
\EndOfBibitem
\bibitem[Fang \emph{et~al.}(2018)Fang, Adjokatse, Shao, Even, and Loi]{Fang}
H.~Fang, S.~Adjokatse, S.~Shao, J.~Even and M.~Loi, \emph{Nature
  Communications}, 2018, \textbf{9}, 243\relax
\mciteBstWouldAddEndPuncttrue
\mciteSetBstMidEndSepPunct{\mcitedefaultmidpunct}
{\mcitedefaultendpunct}{\mcitedefaultseppunct}\relax
\EndOfBibitem
\bibitem[König \emph{et~al.}(2010)König, Casalenuovo, Takeda, Conibeer,
  Guillemoles, Patterson, Huang, and Green]{Koning_2010}
D.~König, K.~Casalenuovo, Y.~Takeda, G.~Conibeer, J.~Guillemoles,
  R.~Patterson, L.~Huang and M.~Green, \emph{Physica E: Low-dimensional Systems
  and Nanostructures}, 2010, \textbf{42}, 2862--2866\relax
\mciteBstWouldAddEndPuncttrue
\mciteSetBstMidEndSepPunct{\mcitedefaultmidpunct}
{\mcitedefaultendpunct}{\mcitedefaultseppunct}\relax
\EndOfBibitem
\bibitem[Limpert and Bremner(2015)]{Limpert}
S.~C. Limpert and S.~P. Bremner, \emph{Applied Physics Letters}, 2015,
  \textbf{107}, 073902\relax
\mciteBstWouldAddEndPuncttrue
\mciteSetBstMidEndSepPunct{\mcitedefaultmidpunct}
{\mcitedefaultendpunct}{\mcitedefaultseppunct}\relax
\EndOfBibitem
\bibitem[Conibeer \emph{et~al.}(2015)Conibeer, Shrestha, Huang, Patterson, Xia,
  Feng, Zhang, Gupta, Tayebjee, Smyth, Liao, Lin, Wang, Dai, and
  Chung]{Conibeer_thermalisation}
G.~Conibeer, S.~Shrestha, S.~Huang, R.~Patterson, H.~Xia, Y.~Feng, P.~Zhang,
  N.~Gupta, M.~Tayebjee, S.~Smyth, Y.~Liao, S.~Lin, P.~Wang, X.~Dai and
  S.~Chung, \emph{Solar Energy Materials and Solar Cells}, 2015, \textbf{135},
  124--129\relax
\mciteBstWouldAddEndPuncttrue
\mciteSetBstMidEndSepPunct{\mcitedefaultmidpunct}
{\mcitedefaultendpunct}{\mcitedefaultseppunct}\relax
\EndOfBibitem
\bibitem[Richter \emph{et~al.}(2017)Richter, Branchi, Valduga~de
  Almeida~Camargo, Zhao, Friend, Cerullo, and Deschler]{Richter}
J.~M. Richter, F.~Branchi, F.~Valduga~de Almeida~Camargo, B.~Zhao, R.~H.
  Friend, G.~Cerullo and F.~Deschler, \emph{Nature communications}, 2017,
  \textbf{8}, 376\relax
\mciteBstWouldAddEndPuncttrue
\mciteSetBstMidEndSepPunct{\mcitedefaultmidpunct}
{\mcitedefaultendpunct}{\mcitedefaultseppunct}\relax
\EndOfBibitem
\bibitem[Rota \emph{et~al.}(1993)Rota, Lugli, Elsaesser, and Shah]{Rota}
L.~Rota, P.~Lugli, T.~Elsaesser and J.~Shah, \emph{Physical Review B}, 1993,
  \textbf{47}, 4226--4237\relax
\mciteBstWouldAddEndPuncttrue
\mciteSetBstMidEndSepPunct{\mcitedefaultmidpunct}
{\mcitedefaultendpunct}{\mcitedefaultseppunct}\relax
\EndOfBibitem
\bibitem[Fröhlich and Mott(1939)]{Frohlich}
H.~Fröhlich and N.~F. Mott, \emph{Proceedings of the Royal Society of London.
  Series A. Mathematical and Physical Sciences}, 1939, \textbf{172},
  94--106\relax
\mciteBstWouldAddEndPuncttrue
\mciteSetBstMidEndSepPunct{\mcitedefaultmidpunct}
{\mcitedefaultendpunct}{\mcitedefaultseppunct}\relax
\EndOfBibitem
\bibitem[Wang \emph{et~al.}(2020)Wang, Jin, Hidalgo, Chu, Snaider, Deng, Zhu,
  Lai, Prezhdo, Correa-Baena, and Huang]{Science-hotcarriers}
T.~Wang, L.~Jin, J.~Hidalgo, W.~Chu, J.~M. Snaider, S.~Deng, T.~Zhu, B.~Lai,
  O.~Prezhdo, J.-P. Correa-Baena and L.~Huang, \emph{Science Advances}, 2020,
  \textbf{6}, eabb1336\relax
\mciteBstWouldAddEndPuncttrue
\mciteSetBstMidEndSepPunct{\mcitedefaultmidpunct}
{\mcitedefaultendpunct}{\mcitedefaultseppunct}\relax
\EndOfBibitem
\bibitem[Takeda \emph{et~al.}(2010)Takeda, Motohiro, K{\"o}nig, Aliberti, Feng,
  Shrestha, and Conibeer]{Conibeer_2010}
Y.~Takeda, T.~Motohiro, D.~K{\"o}nig, P.~Aliberti, Y.~Feng, S.~Shrestha and
  G.~Conibeer, \emph{Applied Physics Express}, 2010, \textbf{3}, 104301\relax
\mciteBstWouldAddEndPuncttrue
\mciteSetBstMidEndSepPunct{\mcitedefaultmidpunct}
{\mcitedefaultendpunct}{\mcitedefaultseppunct}\relax
\EndOfBibitem
\bibitem[Ulatowski \emph{et~al.}(2021)Ulatowski, Farrar, Snaith, Johnston, and
  Herz]{Ulatowski}
A.~M. Ulatowski, M.~D. Farrar, H.~J. Snaith, M.~B. Johnston and L.~M. Herz,
  \emph{ACS Photonics}, 2021, \textbf{8}, 2509--2518\relax
\mciteBstWouldAddEndPuncttrue
\mciteSetBstMidEndSepPunct{\mcitedefaultmidpunct}
{\mcitedefaultendpunct}{\mcitedefaultseppunct}\relax
\EndOfBibitem
\bibitem[Takahashi \emph{et~al.}(2013)Takahashi, Hasegawa, Takahashi, and
  Inabe]{tin1}
Y.~Takahashi, H.~Hasegawa, Y.~Takahashi and T.~Inabe, \emph{Journal of Solid
  State Chemistry}, 2013, \textbf{205}, 39--43\relax
\mciteBstWouldAddEndPuncttrue
\mciteSetBstMidEndSepPunct{\mcitedefaultmidpunct}
{\mcitedefaultendpunct}{\mcitedefaultseppunct}\relax
\EndOfBibitem
\bibitem[Leijtens \emph{et~al.}(2017)Leijtens, Prasanna, Gold-Parker, Toney,
  and McGehee]{tin2}
T.~Leijtens, R.~Prasanna, A.~Gold-Parker, M.~F. Toney and M.~D. McGehee,
  \emph{ACS Energy Letters}, 2017, \textbf{2}, 2159--2165\relax
\mciteBstWouldAddEndPuncttrue
\mciteSetBstMidEndSepPunct{\mcitedefaultmidpunct}
{\mcitedefaultendpunct}{\mcitedefaultseppunct}\relax
\EndOfBibitem
\bibitem[Kahmann \emph{et~al.}(2019)Kahmann, Shao, and Loi]{Kahmann_purity}
S.~Kahmann, S.~Shao and M.~A. Loi, \emph{Advanced Functional Materials}, 2019,
  \textbf{29}, 1902963\relax
\mciteBstWouldAddEndPuncttrue
\mciteSetBstMidEndSepPunct{\mcitedefaultmidpunct}
{\mcitedefaultendpunct}{\mcitedefaultseppunct}\relax
\EndOfBibitem
\bibitem[Levinshtein \emph{et~al.}(1999)Levinshtein, Rumyantsev, Shur, and
  Scientific]{GaAs-diel}
M.~Levinshtein, S.~Rumyantsev, M.~Shur and W.~Scientific, \emph{Handbook Series
  on Semiconductor Parameters: Ternary and quaternary III-V compounds}, World
  Scientific Publishing Company, 1999\relax
\mciteBstWouldAddEndPuncttrue
\mciteSetBstMidEndSepPunct{\mcitedefaultmidpunct}
{\mcitedefaultendpunct}{\mcitedefaultseppunct}\relax
\EndOfBibitem
\bibitem[Raymond \emph{et~al.}(2001)Raymond, Robert, and Bernard]{GaAs-mass}
A.~Raymond, J.~Robert and C.~Bernard, \emph{Journal of Physics C: Solid State
  Physics}, 2001, \textbf{12}, 2289\relax
\mciteBstWouldAddEndPuncttrue
\mciteSetBstMidEndSepPunct{\mcitedefaultmidpunct}
{\mcitedefaultendpunct}{\mcitedefaultseppunct}\relax
\EndOfBibitem
\bibitem[Handa \emph{et~al.}(2020)Handa, Yamada, Nagai, and
  Kanemitsu]{GaAs-freq}
T.~Handa, T.~Yamada, M.~Nagai and Y.~Kanemitsu, \emph{Physical Chemistry
  Chemical Physics}, 2020, \textbf{22}, 26069--26087\relax
\mciteBstWouldAddEndPuncttrue
\mciteSetBstMidEndSepPunct{\mcitedefaultmidpunct}
{\mcitedefaultendpunct}{\mcitedefaultseppunct}\relax
\EndOfBibitem
\bibitem[Brivio \emph{et~al.}(2013)Brivio, Walker, and Walsh]{Brivio_2013}
F.~Brivio, A.~B. Walker and A.~Walsh, \emph{{APL} Materials}, 2013, \textbf{1},
  042111\relax
\mciteBstWouldAddEndPuncttrue
\mciteSetBstMidEndSepPunct{\mcitedefaultmidpunct}
{\mcitedefaultendpunct}{\mcitedefaultseppunct}\relax
\EndOfBibitem
\bibitem[Brivio \emph{et~al.}(2014)Brivio, Butler, Walsh, and van
  Schilfgaarde]{effmassMAPbI3}
F.~Brivio, K.~T. Butler, A.~Walsh and M.~van Schilfgaarde, \emph{Physical
  Review B}, 2014, \textbf{89}, 155204\relax
\mciteBstWouldAddEndPuncttrue
\mciteSetBstMidEndSepPunct{\mcitedefaultmidpunct}
{\mcitedefaultendpunct}{\mcitedefaultseppunct}\relax
\EndOfBibitem
\bibitem[Zhao \emph{et~al.}(2017)Zhao, Skelton, Hu, La-o vorakiat, Zhu, Marcus,
  Michel-Beyerle, Lam, Walsh, and Chia]{Zhao}
D.~Zhao, J.~Skelton, H.~Hu, C.~La-o vorakiat, J.-X. Zhu, R.~Marcus,
  M.~Michel-Beyerle, Y.-M. Lam, A.~Walsh and E.~Chia, \emph{Applied Physics
  Letters}, 2017, \textbf{111}, 201903\relax
\mciteBstWouldAddEndPuncttrue
\mciteSetBstMidEndSepPunct{\mcitedefaultmidpunct}
{\mcitedefaultendpunct}{\mcitedefaultseppunct}\relax
\EndOfBibitem
\bibitem[Mosconi \emph{et~al.}(2015)Mosconi, Umari, and De~Angelis]{Filippo}
E.~Mosconi, P.~Umari and F.~De~Angelis, \emph{J. Mater. Chem. A}, 2015,
  \textbf{3}, 9208--9215\relax
\mciteBstWouldAddEndPuncttrue
\mciteSetBstMidEndSepPunct{\mcitedefaultmidpunct}
{\mcitedefaultendpunct}{\mcitedefaultseppunct}\relax
\EndOfBibitem
\bibitem[Ponce \emph{et~al.}(2019)Ponce, Schlipf, and Giustino]{Ponce}
S.~Ponce, M.~Schlipf and F.~Giustino, \emph{ACS Energy Letters}, 2019,
  \textbf{4}, 456--463\relax
\mciteBstWouldAddEndPuncttrue
\mciteSetBstMidEndSepPunct{\mcitedefaultmidpunct}
{\mcitedefaultendpunct}{\mcitedefaultseppunct}\relax
\EndOfBibitem
\bibitem[Ma \emph{et~al.}(2017)Ma, Li, Li, Lin, Wang, and Qiao]{FaPbI3-eps}
F.~Ma, J.~Li, W.~Li, N.~Lin, L.~Wang and J.~Qiao, \emph{Chemical Science},
  2017, \textbf{8}, 800--805\relax
\mciteBstWouldAddEndPuncttrue
\mciteSetBstMidEndSepPunct{\mcitedefaultmidpunct}
{\mcitedefaultendpunct}{\mcitedefaultseppunct}\relax
\EndOfBibitem
\bibitem[Galkowski \emph{et~al.}(2016)Galkowski, Mitioglu, Miyata, Plochocka,
  Portugall, Eperon, Wang, Stergiopoulos, Stranks, Snaith, and
  Nicholas]{FaPbI3-effmass}
K.~Galkowski, A.~Mitioglu, A.~Miyata, P.~Plochocka, O.~Portugall, G.~E. Eperon,
  J.~T.-W. Wang, T.~Stergiopoulos, S.~D. Stranks, H.~J. Snaith and R.~J.
  Nicholas, \emph{Energy \& Environmental Science}, 2016, \textbf{9},
  962--970\relax
\mciteBstWouldAddEndPuncttrue
\mciteSetBstMidEndSepPunct{\mcitedefaultmidpunct}
{\mcitedefaultendpunct}{\mcitedefaultseppunct}\relax
\EndOfBibitem
\bibitem[Wright \emph{et~al.}(2016)Wright, Verdi, Milot, Eperon, Perez-Osorio,
  Snaith, Giustino, Johnston, and Herz]{FAPbI3-freq}
A.~D. Wright, C.~Verdi, R.~L. Milot, G.~E. Eperon, M.~A. Perez-Osorio, H.~J.
  Snaith, F.~Giustino, M.~B. Johnston and L.~M. Herz, \emph{Nature
  Communications}, 2016, \textbf{7}, 11755\relax
\mciteBstWouldAddEndPuncttrue
\mciteSetBstMidEndSepPunct{\mcitedefaultmidpunct}
{\mcitedefaultendpunct}{\mcitedefaultseppunct}\relax
\EndOfBibitem
\bibitem[Filippetti \emph{et~al.}(2021)Filippetti, Kahmann, Caddeo, Mattoni,
  Saba, Bosin, and Loi]{FaSnI3-diel}
A.~Filippetti, S.~Kahmann, C.~Caddeo, A.~Mattoni, M.~Saba, A.~Bosin and M.~Loi,
  \emph{Journal of Materials Chemistry A}, 2021, \textbf{9}, 11812--11826\relax
\mciteBstWouldAddEndPuncttrue
\mciteSetBstMidEndSepPunct{\mcitedefaultmidpunct}
{\mcitedefaultendpunct}{\mcitedefaultseppunct}\relax
\EndOfBibitem
\bibitem[Ferry(2021)]{Ferry}
D.~K. Ferry, \emph{Hot Carriers in Semiconductors}, IOP Publishing, 2021, pp.
  7--1 to 7--49\relax
\mciteBstWouldAddEndPuncttrue
\mciteSetBstMidEndSepPunct{\mcitedefaultmidpunct}
{\mcitedefaultendpunct}{\mcitedefaultseppunct}\relax
\EndOfBibitem
\bibitem[Jacoboni and Lugli(2011)]{Jacoboni}
C.~Jacoboni and P.~Lugli, \emph{The Monte Carlo Method for Semiconductor Device
  Simulation}, Springer Vienna, 2011\relax
\mciteBstWouldAddEndPuncttrue
\mciteSetBstMidEndSepPunct{\mcitedefaultmidpunct}
{\mcitedefaultendpunct}{\mcitedefaultseppunct}\relax
\EndOfBibitem
\bibitem[Jacoboni and Reggiani(1983)]{Jacoboni2}
C.~Jacoboni and L.~Reggiani, \emph{Reviews of Modern Physics}, 1983,
  \textbf{55}, 645--705\relax
\mciteBstWouldAddEndPuncttrue
\mciteSetBstMidEndSepPunct{\mcitedefaultmidpunct}
{\mcitedefaultendpunct}{\mcitedefaultseppunct}\relax
\EndOfBibitem
\bibitem[Hess(1991)]{Hess}
K.~Hess, \emph{Monte Carlo Device Simulation: Full Band and Beyond}, Kluwer
  Academic Publishers, USA, 1991\relax
\mciteBstWouldAddEndPuncttrue
\mciteSetBstMidEndSepPunct{\mcitedefaultmidpunct}
{\mcitedefaultendpunct}{\mcitedefaultseppunct}\relax
\EndOfBibitem
\bibitem[Hockney and Eastwood(1988)]{Hockney}
R.~Hockney and J.~Eastwood, \emph{Computer Simulation Using Particles}, CRC
  Press, 1988\relax
\mciteBstWouldAddEndPuncttrue
\mciteSetBstMidEndSepPunct{\mcitedefaultmidpunct}
{\mcitedefaultendpunct}{\mcitedefaultseppunct}\relax
\EndOfBibitem
\bibitem[Tomizawa(1993)]{Tomizawa}
K.~Tomizawa, \emph{Numerical simulation of submicron semiconductor devices},
  1993\relax
\mciteBstWouldAddEndPuncttrue
\mciteSetBstMidEndSepPunct{\mcitedefaultmidpunct}
{\mcitedefaultendpunct}{\mcitedefaultseppunct}\relax
\EndOfBibitem
\bibitem[Kosina \emph{et~al.}(2000)Kosina, Nedjalkov, and Selberherr]{Kosina}
H.~Kosina, M.~Nedjalkov and S.~Selberherr, \emph{IEEE Transactions on Electron
  Devices}, 2000, \textbf{47}, 1898--1908\relax
\mciteBstWouldAddEndPuncttrue
\mciteSetBstMidEndSepPunct{\mcitedefaultmidpunct}
{\mcitedefaultendpunct}{\mcitedefaultseppunct}\relax
\EndOfBibitem
\bibitem[Osman and Ferry(1987)]{GaAs_EMC}
M.~A. Osman and D.~K. Ferry, \emph{Physical Review B}, 1987, \textbf{36},
  6018--6032\relax
\mciteBstWouldAddEndPuncttrue
\mciteSetBstMidEndSepPunct{\mcitedefaultmidpunct}
{\mcitedefaultendpunct}{\mcitedefaultseppunct}\relax
\EndOfBibitem
\bibitem[Lugli and Ferry(1986)]{Lugli_EMC}
P.~Lugli and D.~Ferry, \emph{Physical review letters}, 1986, \textbf{56},
  1295--1297\relax
\mciteBstWouldAddEndPuncttrue
\mciteSetBstMidEndSepPunct{\mcitedefaultmidpunct}
{\mcitedefaultendpunct}{\mcitedefaultseppunct}\relax
\EndOfBibitem
\bibitem[Duncan \emph{et~al.}(1998)Duncan, Ravaioli, and Jakumeit]{Duncan_EMC}
A.~Duncan, U.~Ravaioli and J.~Jakumeit, \emph{IEEE Transactions on Electron
  Devices}, 1998, \textbf{45}, 867 -- 876\relax
\mciteBstWouldAddEndPuncttrue
\mciteSetBstMidEndSepPunct{\mcitedefaultmidpunct}
{\mcitedefaultendpunct}{\mcitedefaultseppunct}\relax
\EndOfBibitem
\bibitem[Irvine \emph{et~al.}(2021)Irvine, Walker, and Wolf]{Walker}
L.~A.~D. Irvine, A.~B. Walker and M.~J. Wolf, \emph{Physical Review B}, 2021,
  \textbf{103}, L220305\relax
\mciteBstWouldAddEndPuncttrue
\mciteSetBstMidEndSepPunct{\mcitedefaultmidpunct}
{\mcitedefaultendpunct}{\mcitedefaultseppunct}\relax
\EndOfBibitem
\bibitem[Gollner \emph{et~al.}(2023)Gollner, Steiner, and
  Filipovic]{vienna-EMC}
L.~Gollner, R.~Steiner and L.~Filipovic, \emph{Vienna-EMC},
  \url{https://github.com/ViennaTools/ViennaEMC}, 2023\relax
\mciteBstWouldAddEndPuncttrue
\mciteSetBstMidEndSepPunct{\mcitedefaultmidpunct}
{\mcitedefaultendpunct}{\mcitedefaultseppunct}\relax
\EndOfBibitem
\bibitem[Nederveen(1989)]{phd-EMC}
K.~Nederveen, \emph{PhD thesis}, Electrical Engineering, 1989\relax
\mciteBstWouldAddEndPuncttrue
\mciteSetBstMidEndSepPunct{\mcitedefaultmidpunct}
{\mcitedefaultendpunct}{\mcitedefaultseppunct}\relax
\EndOfBibitem
\bibitem[L’Ecuyer(1990)]{Ecuyer}
P.~L’Ecuyer, \emph{Communications of the ACM}, 1990, \textbf{33},
  85--97\relax
\mciteBstWouldAddEndPuncttrue
\mciteSetBstMidEndSepPunct{\mcitedefaultmidpunct}
{\mcitedefaultendpunct}{\mcitedefaultseppunct}\relax
\EndOfBibitem
\bibitem[Blanchard \emph{et~al.}(2015)Blanchard, Bramas, Coulaud, Darve, Dupuy,
  Etcheverry, and Sylvand]{ScalFMM}
P.~Blanchard, B.~Bramas, O.~Coulaud, E.~F. Darve, L.~Dupuy, A.~Etcheverry and
  G.~Sylvand, IEEE International Conference on Computational Science and
  Engineering, 2015\relax
\mciteBstWouldAddEndPuncttrue
\mciteSetBstMidEndSepPunct{\mcitedefaultmidpunct}
{\mcitedefaultendpunct}{\mcitedefaultseppunct}\relax
\EndOfBibitem
\bibitem[Rokhlin(1985)]{rokhlin}
V.~Rokhlin, \emph{Journal of Computational Physics}, 1985, \textbf{60}, 187 --
  207\relax
\mciteBstWouldAddEndPuncttrue
\mciteSetBstMidEndSepPunct{\mcitedefaultmidpunct}
{\mcitedefaultendpunct}{\mcitedefaultseppunct}\relax
\EndOfBibitem
\bibitem[Heitzinger \emph{et~al.}(2004)Heitzinger, Ringhofer, Ahmed, and
  Vasileska]{vasileska}
C.~Heitzinger, C.~Ringhofer, S.~Ahmed and D.~Vasileska, \emph{Journal of
  Computational Electronics}, 2004,  24 -- 25\relax
\mciteBstWouldAddEndPuncttrue
\mciteSetBstMidEndSepPunct{\mcitedefaultmidpunct}
{\mcitedefaultendpunct}{\mcitedefaultseppunct}\relax
\EndOfBibitem
\bibitem[P{\"o}tz(1987)]{hotphbottle}
W.~P{\"o}tz, \emph{Physical Review B}, 1987, \textbf{36}, 5016\relax
\mciteBstWouldAddEndPuncttrue
\mciteSetBstMidEndSepPunct{\mcitedefaultmidpunct}
{\mcitedefaultendpunct}{\mcitedefaultseppunct}\relax
\EndOfBibitem
\bibitem[Prabhu and Vengurlekar(1996)]{hotphbottle2}
S.~S. Prabhu and A.~S. Vengurlekar, \emph{Physical Review B}, 1996,
  \textbf{53}, 7815--7818\relax
\mciteBstWouldAddEndPuncttrue
\mciteSetBstMidEndSepPunct{\mcitedefaultmidpunct}
{\mcitedefaultendpunct}{\mcitedefaultseppunct}\relax
\EndOfBibitem
\bibitem[Del~Fatti \emph{et~al.}(2000)Del~Fatti, Voisin, Achermann, Tzortzakis,
  Christofilos, and Vall\'ee]{delfatti}
N.~Del~Fatti, C.~Voisin, M.~Achermann, S.~Tzortzakis, D.~Christofilos and
  F.~Vall\'ee, \emph{Physical Review B}, 2000, \textbf{61}, 16956--16966\relax
\mciteBstWouldAddEndPuncttrue
\mciteSetBstMidEndSepPunct{\mcitedefaultmidpunct}
{\mcitedefaultendpunct}{\mcitedefaultseppunct}\relax
\EndOfBibitem
\bibitem[Zhu \emph{et~al.}(2016)Zhu, Miyata, Fu, Wang, Joshi, Niesner,
  Williams, Jin, and Zhu]{Zhu_polarons}
H.~Zhu, K.~Miyata, Y.~Fu, J.~Wang, P.~P. Joshi, D.~Niesner, K.~W. Williams,
  S.~Jin and X.-Y. Zhu, \emph{Science}, 2016, \textbf{353}, 1409--1413\relax
\mciteBstWouldAddEndPuncttrue
\mciteSetBstMidEndSepPunct{\mcitedefaultmidpunct}
{\mcitedefaultendpunct}{\mcitedefaultseppunct}\relax
\EndOfBibitem
\bibitem[Frost \emph{et~al.}(2017)Frost, Whalley, and Walsh]{Frost_hotpolarons}
J.~M. Frost, L.~D. Whalley and A.~Walsh, \emph{ACS Energy Letters}, 2017,
  \textbf{2}, 2647--2652\relax
\mciteBstWouldAddEndPuncttrue
\mciteSetBstMidEndSepPunct{\mcitedefaultmidpunct}
{\mcitedefaultendpunct}{\mcitedefaultseppunct}\relax
\EndOfBibitem
\bibitem[Evans \emph{et~al.}(2018)Evans, Miyata, Joshi, Maehrlein, Liu, and
  Zhu]{Evans_polarons}
T.~J.~S. Evans, K.~Miyata, P.~P. Joshi, S.~Maehrlein, F.~Liu and X.-Y. Zhu,
  \emph{The Journal of Physical Chemistry C}, 2018, \textbf{122},
  13724--13730\relax
\mciteBstWouldAddEndPuncttrue
\mciteSetBstMidEndSepPunct{\mcitedefaultmidpunct}
{\mcitedefaultendpunct}{\mcitedefaultseppunct}\relax
\EndOfBibitem
\bibitem[Joshi \emph{et~al.}(2019)Joshi, Maehrlein, and Zhu]{Joshi_polarons}
P.~P. Joshi, S.~F. Maehrlein and X.~Zhu, \emph{Advanced Materials}, 2019,
  \textbf{31}, 1803054\relax
\mciteBstWouldAddEndPuncttrue
\mciteSetBstMidEndSepPunct{\mcitedefaultmidpunct}
{\mcitedefaultendpunct}{\mcitedefaultseppunct}\relax
\EndOfBibitem
\bibitem[Burgos-Caminal \emph{et~al.}(2021)Burgos-Caminal, Moreno-Naranjo,
  Willauer, Paraecattil, Ajdarzadeh, and Moser]{Burgos-Caminal}
A.~Burgos-Caminal, J.~M. Moreno-Naranjo, A.~R. Willauer, A.~A. Paraecattil,
  A.~Ajdarzadeh and J.-E. Moser, \emph{The Journal of Physical Chemistry C},
  2021, \textbf{125}, 98--106\relax
\mciteBstWouldAddEndPuncttrue
\mciteSetBstMidEndSepPunct{\mcitedefaultmidpunct}
{\mcitedefaultendpunct}{\mcitedefaultseppunct}\relax
\EndOfBibitem
\bibitem[Emin(2012)]{Emin}
D.~Emin, \emph{Polarons}, Cambridge University Press, 2012\relax
\mciteBstWouldAddEndPuncttrue
\mciteSetBstMidEndSepPunct{\mcitedefaultmidpunct}
{\mcitedefaultendpunct}{\mcitedefaultseppunct}\relax
\EndOfBibitem
\end{mcitethebibliography}
\bibliographystyle{main} 

\end{document}